\documentclass[11pt]{article}
\pdfoutput=1

\usepackage{euscript}
\usepackage{amssymb}
\usepackage{amsfonts}
\usepackage{amsbsy}
\usepackage{epsfig}
\usepackage{amsthm}
\usepackage{amscd}
\usepackage{amstext}
\usepackage{verbatim}
\usepackage{amsmath}

\def\ben{\begin{equation}}
\def\een{\end{equation}}
\def\half{{\textstyle{\frac12}}}

\let\a=\alpha \let\b=\beta \let\g=\gamma \let\d=\delta 
   
     \let\r=\rho

\let\w=\omega

\newcommand{\ba}{\begin{align}}
\newcommand{\ea}{\end{align}}
\newcommand{\nn}{\nonumber \\}
\newcommand{\bi}{\begin{itemize}}
\newcommand{\ei}{\end{itemize}}

\def\be{\begin{equation}}
\def\ee{\end{equation}}
\def\beq{\begin{equation}}
\def\eeq{\end{equation}}

\def\dalemb#1#2{{\vbox{\hrule height .#2pt
       \hbox{\vrule width.#2pt height#1pt \kern#1pt
               \vrule width.#2pt}
       \hrule height.#2pt}}}

\newcommand{\bea}{\begin{eqnarray}}
\newcommand{\eea}{\end{eqnarray}}

\def\R{{{\Bbb R}}}

\def\ocal{{\mathcal{O}}}

\begin{document}

\makeatletter
\renewcommand{\theequation}{\thesection.\arabic{equation}}
\@addtoreset{equation}{section}
\makeatother

\begin{flushright}
\end{flushright}

\begin{center}

{ \Large {\bf Universal linear in temperature resistivity from \\
black hole superradiance
}}

\vspace{1cm}

Aristomenis Donos$^\sharp$ and Sean A. Hartnoll$^\flat$
\vspace{0.7cm}

{\small
{\it $^\sharp$ Blackett Laboratory, 
        Imperial College, \\ London, SW7 2AZ, U.K. \\
        \vspace{0.3cm}
       $^\flat$  Department of Physics, Stanford University, \\
Stanford, CA 94305-4060, USA }}

\vspace{1.6cm}

\end{center}

\begin{abstract}

Observations across many families of unconventional materials motivative the search for robust mechanisms producing linear in temperature d.c.$\,$resistivity. BKT quantum phase transitions are commonplace in holographic descriptions of finite density matter, separating critical and ordered phases. We show that at a holographic BKT critical point, if the unstable operator is coupled to the current via irrelevant operators, then a linear contribution to the resistivity is universally obtained. We also obtain broad power law tails in the optical conductivity, that shift spectral weight from the Drude peak as well as interband energy scales. We give a partial realization of this scenario using an Einstein-Maxwell-pseudoscalar bulk theory. The instability is a vectorial mode at nonzero wavevector, which is communicated to the homogeneous current via irrelevant coupling to an ionic lattice.

\end{abstract}

\pagebreak
\setcounter{page}{1}
\setcounter{equation}{0}


\section{Unconventional d.c. and optical conductivities}
\label{sec:intro}

It is an observed fact that several chemically diverse families of unconventional materials exhibit a low temperature resistivity that is linear in temperature, when tuned to what appear to be quantum critical points.
Illustrative recent experimental data can be found for cuprates, pnictides, heavy fermions, ruthenates and organic superconductors -- see \cite{Sachdev:2011cs} for an overview and references. While it is likely that there may be more than one explanation for this behavior, the universality of the temperature dependence of the resistivity seen in the phase diagrams of these materials strongly motivates the search for robust physical mechanisms that can reproduce the observations.

In this paper we show, in the context of the holographic correspondence, that the structure of superradiant instabilities of extremal black hole horizons leads universally to a linear in temperature resistivity when these systems are tuned to the quantum critical point mediating the instability. This mechanism will be described in some generality. We go on to illustrate the process in a particular model that captures additional features of several of the unconventional materials of interest: the resistivity will be due to scattering off modes that are becoming unstable, and which are supported at nonzero momentum. One can take these modes to model the spin density and charge density wave instabilities appearing in the phase diagrams of \cite{Sachdev:2011cs}.

That the resistivity is linear rather than, say, quadratic in temperature is only half of the mystery. Unlike conventional metals, many of the materials of interest fail to exhibit resistivity saturation as the temperature is increased. The resistivity increases unabated with temperature through the Mott-Ioffe-Regel limit. The materials are consequently known as bad metals \cite{Emery:1995zz}. This fact becomes particularly confusing when considered in conjunction with the dependence of the in-plane (frequency dependent) optical conductivity on temperature. In conventional metals, the optical conductivity exhibits a Drude peak that broadens as the temperature is increased. Resistivity saturation occurs when the spectral weight is smeared out over the whole of the bandwidth and the Drude peak effectively disappears \cite{hussey}. In several classes of unconventional materials that do not exhibit saturation, the Drude peak is accompanied by an extended tail that falls off slowly at large frequencies up to the bandwidth scale \cite{basov}. At the would-be saturation temperature, the Drude peak is observed to melt into this broader feature \cite{hussey, basov}. The d.c.$\,$resistivity, however, continues to increase linearly in temperature with the same slope, irrespectively of whether or not there is an associated Drude peak! The conundrum is served: Does the linear in temperature resistivity originate in Drude-like (i.e. momentum-relaxing) scattering or not?

The answer to this question is again likely to be material dependent. The picture of resistivity saturation outline above may not be correct. Nonetheless, the characterization of bad metals as metals that do not exhibit a zero frequency collective mode encourages the notion that the resistivity might be controlled by quantum critical physics, presumably responsible for the extended tail in the spectral density, rather than be sensitive to the mechanism of momentum relaxation. Such a picture is likely to have trouble with the fact that at lower temperatures a very sharp Drude peak is observed on top of the broader feature, see e.g. \cite{boris} for measurements in optimally doped YBCO. The mechanism of linear resistivity presented in this paper will be quantum critical in nature, and we will assume that the Drude peak has been swamped by the critical degrees of freedom.

Figure \ref{fig:optical} below illustrates the above discussion.

\begin{figure}[h]
\begin{center}
\includegraphics[width = \textwidth]{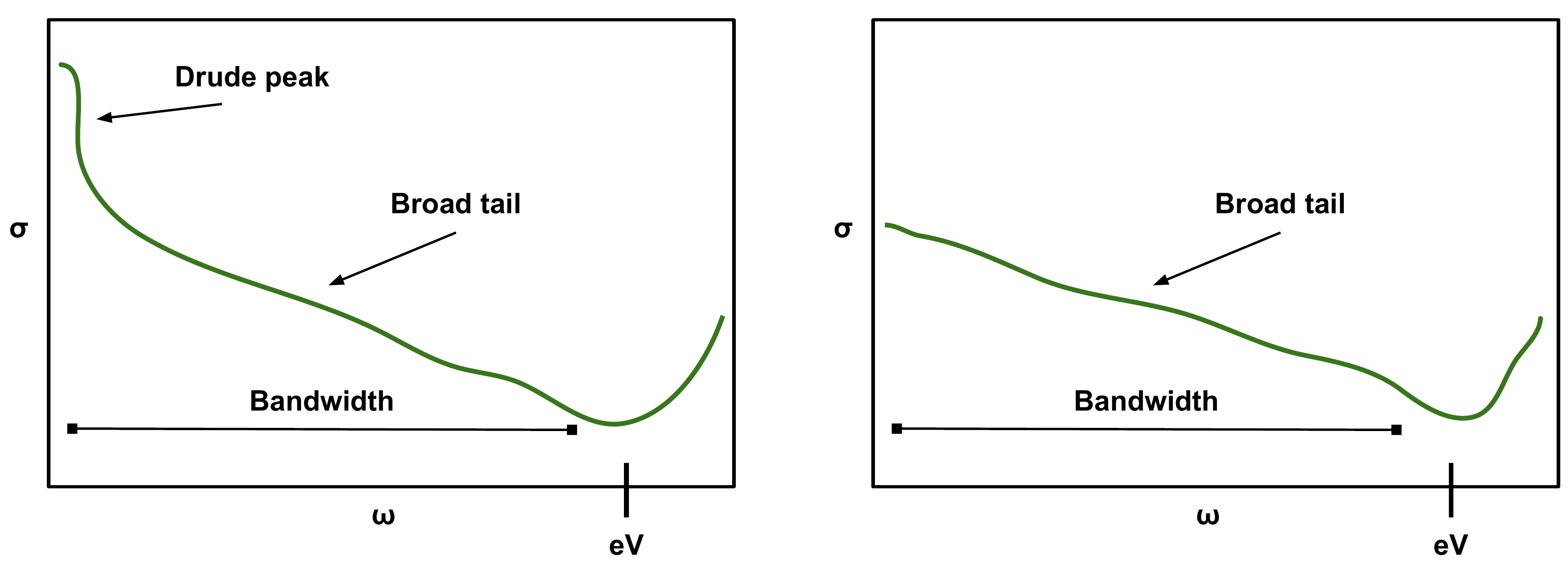}\caption{Theorist's schematic view of the optical conductivity in bad metals at lower temperatures (left) and higher temperatures (right). As the temperature is raised, spectral weight is shifted from the Drude peak into the broad tail and to interband energy scales. The linear in temperature d.c.$\,$resistivity does not notice the melting of the Drude peak.
\label{fig:optical}}
\end{center}
\end{figure}

\section{Extremal horizons and BKT transitions}
\label{sec:bkt}

In quantum theories with a holographically dual description, the strongly interacting physics is described by classical gravitational dynamics in a dual spacetime with one extra spatial dimension. The extra dimension geometrically implements the renormalization group flow. In particular, the far interior of the spacetime describes the far IR, or lowest energy scales, of the dual quantum system \cite{Hartnoll:2011fn}. In many holographic models placed at a finite charge density, $\langle J^t \rangle \neq 0$, the far IR geometry is characterized by an emergent scaling symmetry that indicates a power law specific heat at low temperatures, and hence the presence of gapless degrees of freedom \cite{Hartnoll:2011fn}. Working in two spatial dimensions for concreteness, the general scale invariant form of the metric is \cite{Kachru:2008yh}
\be\label{eq:scalingIR}
ds^2_{\text{IR}} = L^2_{\text{IR}} \left(\frac{- dt^2 + d\r^2}{\r^2} + \frac{dx^2 + dy^2}{\r^{2/z}} \right) \,.
\ee
This spacetime admits the scaling symmetry $\{t,\r\} \to \lambda \{t,\r\}, \vec x \to \lambda^{1/z} \vec x$. Therefore $z$ is the dynamical critical exponent. For simplicity we will not discuss here the more general class of metrics describing hyperscaling violation \cite{Charmousis:2010zz, Huijse:2011ef}, although our considerations apply to those spacetimes also. Holographic IR scaling spacetimes typically arise over a finite range of parameter space and therefore describe critical phases.

In a scaling geometry of the form (\ref{eq:scalingIR}), all the operators $\ocal$ in the low energy theory have an energy scaling dimension $\Delta$. This dimension determines the spectral weight (imaginary part of the retarded Green's function) in the regime $\w \ll T \ll \mu$ to be
\be\label{eq:im}
\lim_{\w \to 0}\frac{1}{\w} \text{Im} \, G_{\ocal \ocal}^R(\w,T) \sim T^{2 \Delta - 2 - 2/z}\,.
\ee
Here the chemical potential $\mu$ has been used to indicate the UV scale. Physics well below that scale is captured by the low energy spacetime (\ref{eq:scalingIR}) and is amenable to dimensional analysis. We also used the fact that the spectral weight is odd in frequency for bosonic operators. In general, the above spectral weight is evaluated at zero momentum, $k=0$. A slight generalization is possible in the case of $z = \infty$. In this case, space does not scale and so the momentum $k$ is dimensionless and can be nonzero. The scaling dimension becomes
momentum dependent: $\Delta(k)$.

According to the basic holographic dictionary \cite{Witten:1998qj, Gubser:1998bc}, each operator $\ocal$ is dual to a field $\phi$ in the IR spacetime. For simplicity, to start with, consider the case in which $\phi$ is a scalar field with mass $m$ that is not coupled to other fields at a linearized level. By solving the bulk wave equation in the geometry (\ref{eq:scalingIR}) and reading off the scaling behavior from the solution as $\rho \to 0$, one
immediately finds
\be\label{eq:delta}
\Delta =  \frac{2 + z}{2 z} + \nu \equiv \frac{2 + z}{2 z} + \sqrt{L^2_\text{IR} m^2 + \frac{(2+z)^2}{4 z^2}} \,.
\ee
The parameter $\nu$ introduced here will appear repeatedly below.
In the case of $z = \infty$, the mass $m^2$ can be momentum dependent. In this expression we see that if the mass squared satisfies the generalized Breitenlohner-Freedman bound
\be\label{eq:bf}
L^2_\text{IR} m^2 > -  \frac{(2+z)^2}{4 z^2} \,,
\ee
then the scaling dimension $\Delta$ is real.

The quantity $L^2_\text{IR} m^2$ can be tuned by varying UV parameters. In particular, we can imagine tuning these parameters such that the mass squared drops below the bound (\ref{eq:bf}), and the scaling dimension becomes complex. It is well understood by now that this triggers an interesting quantum phase transition in which the operator $\ocal$ condenses once the mass squared of $\phi$ becomes too negative. We will return to the nature of the transition very shortly, but first we notice that precisely at the critical point, where the square root in (\ref{eq:delta}) vanishes and $\nu = 0$, then from (\ref{eq:im}) and (\ref{eq:delta}) we have
\be\label{eq:critical}
\lim_{\w \to 0}\frac{1}{\w} \text{Im} \, G_{\ocal \ocal}^R(\w) \sim \frac{1}{T} \,.
\ee
This expression, which has been previously emphasized in \cite{Iqbal:2011aj, Vegh:2011aa}, will be at the heart of the linear resistivity above a quantum critical point that we will discuss in the following section. Previous interest in this expression is due to the fact that it resembles the spectral weight of the bosonic mode underlying the marginal Fermi liquid phenomenology of the cuprates \cite{Varma:1989zz}. It is not quite the same, however, because in general (\ref{eq:critical}) only holds at $k=0$. Even in the case of $z = \infty$, because the dimension $\Delta$ is $k$ dependent -- this is why these systems were termed semi-locally critical in \cite{Iqbal:2011in}, rather than fully locally critical -- the full spectral density will have a nontrivial $k$ dependence, and (\ref{eq:critical}) will only hold for one value of $k$ at a time. The absence of a $k$ dependence is crucial for the marginal Fermi liquid, as the mode is coupled to a Fermi surface, which has spectral weight at a nonzero $k=k_F$ that is set by UV dynamics. In contrast, our framework will operate entirely at the level of currents, which control the holographic charge dynamics at leading classical order in the bulk, and will not involve explicit discussion of Fermi surfaces.

The association of complex IR scaling dimensions to instabilities was first made in the context of holographic superconductors \cite{Hartnoll:2008kx, Denef:2009tp, Gubser:2008pf}. In cases where the operator $\ocal$ carries a charge, the instability can be understood as a cousin of the superradiant instabilities of charged black holes, driven by pair production of quanta near the horizon and leading to a discharging of the black hole \cite{Hartnoll:2011fn}.
It was later realized that the quantum phase transition mediating this instability was of Berezinskii-Kosterlitz-Thouless (BKT) type.  For instance, when the mass squared is just below the bound (\ref{eq:bf}), the temperature below which the instability occurs scales like
\be\label{eq:tc}
T_c \sim \mu \, e^{- \pi/\sqrt{-\nu^2}} \,.
\ee
Similar exponential hierarchies control quantities such as the condensate just below the critical mass squared.
Such zero temperature BKT transitions were first discussed in \cite{Kaplan:2009kr}.
Unlike the conventional BKT transition, they can occur in any dimension and are not tied to an interpretation of vortex unbinding, but rather describe the merger of a UV and IR fixed point. These transitions were subsequently noted to be rather generic in holographic settings \cite{Jensen:2010ga, Iqbal:2010eh} and to admit a `semi-holographic' description \cite{Jensen:2011af, Iqbal:2011aj}, in which the only role of holography is to provide a critical IR sector in which the transition occurs \cite{Faulkner:2010tq}. Our discussion in the following section will be essentially semi-holographic in nature, being independent of most details of the UV region of the bulk geometry. In \S \ref{sec:broader} of the discussion section we will take a further step back from holography and comment on the validity of the result we have just described for general BKT transitions.

There are two key points we would like the reader to take from the above. Firstly, that given a `critical phase' with an IR scaling symmetry described by the metric (\ref{eq:scalingIR}) one can induce a quantum phase transition by tuning the dimension of an operator to become complex. Secondly, at the corresponding quantum critical point, the spectral density of this operator has the temperature dependence (\ref{eq:critical}). This last statement holds at $k=0$ for finite $z$, and at some specific $k_\star$ when $z = \infty$.

\section{Mechanism of linear resistivity}
\label{sec:mechanism}

The d.c.$\,$conductivity is given by
\be\label{eq:sigma}
\sigma  = \lim_{\w \to 0} \frac{1}{\w} \, \text{Im} \, G^R_{J^x J^x}(\w,T) \,.
\ee
At a nonzero charge density $\langle J^t \rangle$ and if momentum is conserved, this quantity is problematic because, in addition to the contribution (\ref{eq:sigma}), there is a delta function in the dissipative conductivity at $\w = 0$. In order to relax momentum over experimental timescales, the charge carriers must either be parametrically diluted or must interact with parametrically heavier degrees of freedom \cite{Hartnoll:2012rj}. The result is the broadening of the delta function into a Drude peak. The contribution of the Drude peak to the d.c. conductivity is intimately connected to the mechanism by which momentum is relaxed. For instance, umklapp scattering in a Fermi liquid gives rise to the celebrated $T^2$ dependence of the resistivity. Instead, we would like the d.c.$\,$conductivity to be dominated by the universal quantum critical dynamics underlying the spectral weight (\ref{eq:critical}). This can happen if the Drude peak contribution is swamped by the zero frequency limit of an extended tail that arises from scattering off critical modes. As we discussed in \S \ref{sec:intro} above, this may be the case in bad metallic regimes.

It is sometimes asserted that the presence of a Drude peak is synonymous with a quasiparticle description of transport. This is not quite correct. The essential requirement for a Drude peak is a hierarchy of time scales, whereby the momentum relaxation timescale is much longer than any other timescale in the system. This is particularly clear in hydrodynamic or memory matrix approaches, e.g. \cite{Hartnoll:2007ih, Hartnoll:2012rj}. In the presence of such a hierarchy, strongly correlated systems without a quasiparticle description will still exhibit a Drude peak. Conversely, 
the absence of a Drude peak simply requires that momentum is being dumped by the charge carriers at a rate comparable to all other interactions. In such circumstances, the spectral weight from the delta function is transferred into the critical tail or indeed to interband energy scales. Strong interactions are presumably important here to maintain a metallic character and avoid localization \cite{Emery:1995zz}. In the concrete model we consider in the following section, we assume that such a process is occurring in our system, without disrupting the momentum-conserving interactions that we consider, so that in effect we can ignore the delta function contribution to the conductivity. More generally, we can imagine that momentum-relaxing processes are already part of the critical system that is undergoing the quantum critical BKT transition.

If $J^x$ itself were the operator undergoing the BKT transition, then combining (\ref{eq:critical}) and (\ref{eq:sigma}) would directly give a linear in temperature resistivity at the critical point. Such a phase transition would correspond to the spontaneous generation of a uniform current and presumably requires spontaneous symmetry breaking of the global $U(1)$ symmetry. We are not aware of holographic, or other, models where this occurs. However, it is easy for the IR critical behavior (\ref{eq:critical}) to get communicated to the current operator via operators that are irrelevant from the IR quantum critical point of view.

The IR scaling geometry (\ref{eq:scalingIR}) is generically deformed by irrelevant operators that drive a renormalization group flow up towards the finite density UV fixed point or cutoff. In the IR scaling regime, before these irrelevant operators kick in, we can imagine diagonalizing the equations of motion for a generic perturbation of the background to obtain decoupled gauge invariant fields $\Phi_I$. Near the boundary ($\rho \to 0$) of the IR geometry, these satisfy
\be\label{eq:GIR}
\Phi^I(\rho) \sim c^I \Big( \rho^{1+2/z-\Delta_I} + \rho^{\Delta_I} \, {\mathcal G}^R_I(\w,T) \Big) \,.
\ee
Here ${\mathcal G}^R_I(\w,T)$ is the IR Green's function, the $c^I$ are constants, and $\Delta_I$ is the IR dimension of the operator. We have allowed a temperature $T \ll \mu$ that only affects the IR spacetime. One of these dimensions $\Delta_I$ is assumed to undergo a BKT transition of the form described in the previous section.

Away from the IR geometry, the irrelevant operators will typically couple these perturbations. However, in the regime of interest $\w,T \ll \mu$, this coupling does not introduce any additional non-analytic temperature or frequency dependence. In appendix \ref{eq:matching} we show that a generalization of the usual matching procedure of e.g.
\cite{Faulkner:2011tm} implies that to leading order at low frequencies and temperatures\footnote{To be precise, as explained in the appendix, equation (\ref{eq:imgsum}) holds when the $\nu_I$'s are real, that is, on the stable side of the quantum critical point. When some of the $\nu_I$'s are imaginary, the general expression has a complicated logarithmic dependence on $T$ and $\w$.}
\be\label{eq:imgsum}
\text{Im} \, G^R_{J^x J^x}(\w,T) = \sum_I d^I \, \text{Im} \, {\mathcal G}^R_I(\w,T) \,.
\ee
Here the real coefficients $d^I$ will generically be nonzero if the irrelevant operators mix the IR mode $\Phi_I$ with the gauge field $A_x$. This is by no means automatic; the model of the following sections will achieve the required mixing by combining several interesting ingredients. In more familiar condensed matter language, we can think of the direct coupling in (\ref{eq:imgsum}) between the current and a fluctuating order parameter as a cousin of the Aslamazov-Larkin process \cite{al}, in this case mediated by irrelevant operators. The key fact is that we couple the fluctuating field directly to the current, not going via e.g. a fermionic self-energy. It is now immediate that if one of these IR operators undergoes a BKT transition, then the spectral weight (\ref{eq:critical}), plugged into the matching formula (\ref{eq:imgsum}) leads to a linear in temperature resistivity at the critical point
\be\label{eq:linearT}
r = \frac{1}{\sigma} \sim T \,.
\ee
Note that, according to (\ref{eq:im}) and (\ref{eq:delta}), the contribution from a general IR operator $\ocal_I$ to the conductivity is
\be\label{eq:2nu}
\sigma = T^{-1 + 2 \nu_I} \,.
\ee
Recall that $\nu$ was defined in (\ref{eq:delta}). Assuming we are in a stable phase, then we see that the contribution of all the other operators with $\nu_I > 0$ are subleading in the sum (\ref{eq:imgsum}) compared to the critical operator with $\nu = 0$. The above discussion is depicted in figure \ref{fig:phased} below.

\begin{figure}[h]
\begin{center}
\includegraphics[height=270pt]{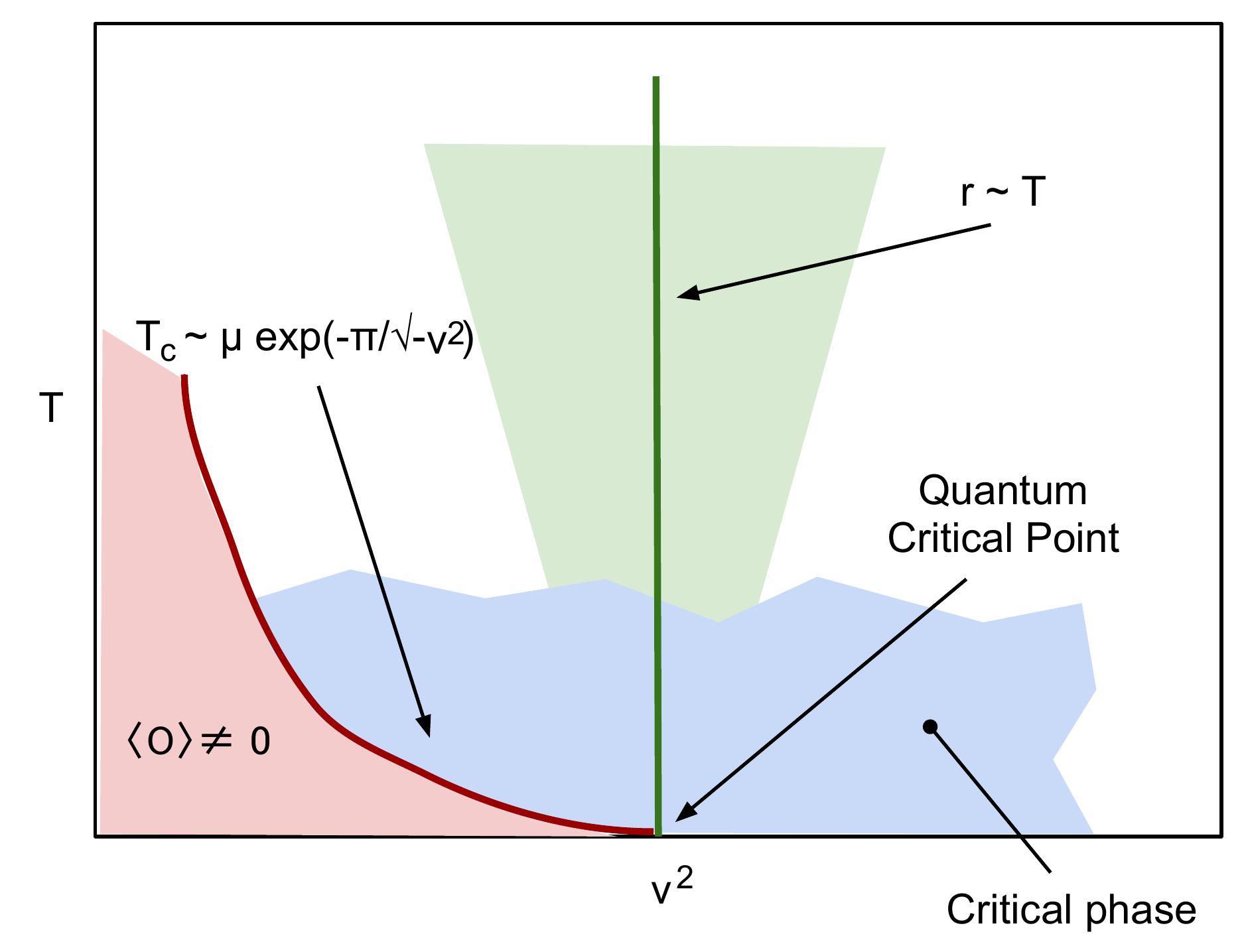}\caption{Schematic phase diagram. The BKT quantum phase transition occurs at the boundary of a quantum critical phase when the scaling dimension of an operator becomes complex, signaling a condensation instability. If the unstable operator is coupled to the current via irrelevant operators, then above the critical point the quantum critical contribution to the resistivity is linear in temperature. Close to the critical point, the mode becoming unstable has a strong effect on the d.c.$\,$and optical conductivities.
\label{fig:phased}}
\end{center}
\end{figure}

While (\ref{eq:2nu}) allows the critical operator to dominate at the critical point, away from the critical point, on the disordered side, it implies that the exponent of the resistivity will be $1 - 2 \nu < 1$. Here $\nu$ is the exponent of the critical operator away from the critical point. This is in contrast to experiments in all the unconventional materials of interest, which show that when detuned from criticality, the exponent of the resistivity increases towards the Fermi liquid $T^2$ behavior \cite{Sachdev:2011cs}. To capture this behavior we would need to increase the scope of our model. The simplest way to do this would be to combine the critical tail contribution we are discussing with a more conventional Femi liquid Drude peak contribution. It may be that by coupling the critical operator to the Fermi liquid, along the marginal Fermi liquid lines of \cite{Iqbal:2011aj, Vegh:2011aa} or otherwise, one can remove the $T^2$ contribution to the resistivity in the critical region of the phase diagram. For instance, very schematically, a form like
\be\label{eq:fudged}
\sigma \sim \nu \, T^{-2} + T^{-1 + 2 \nu} \,,
\ee
would start to look closer to the data.

Complications with the Drude peak are removed if one looks at the optical conductivity. The scaling arguments above now imply that the dissipative conductivity scales like
\be\label{eq:sw}
\sigma \sim \w^{-1 + 2 \nu} \,.
\ee
This gives the marginal Fermi liquid form
\be\label{eq:1w}
\sigma \sim \frac{1}{\w} \,,
\ee
at the critical point. This may be compatible with observations in e.g. YBCO \cite{Varma:1989zz} and LSCO \cite{lsco}. We discuss the optical conductivity in more detail below. It would be interesting to measure in detail the doping dependence of the power law tail of the optical conductivity in these materials, to assess the plausibility of a dependence like (\ref{eq:sw}) close to the critical point. As we noted above, the expressions (\ref{eq:sw}) and (\ref{eq:2nu}) are strictly applicable only on the $\nu > 0$ side of the quantum phase transition. It is a feature of these models that the d.c.$\,$and optical conductivities have asymmetric behavior on opposite sides of the quantum critical point.

There may seem to be a tension in the fact that $J^x$ does not acquire a vacuum expectation value and yet its correlator couples at the lowest frequencies to the unstable mode according to (\ref{eq:imgsum}). We will verify explicitly in our concrete model below that these two statements are compatible.

The mechanism of linear in temperature resistivity we have described is universal in the sense that it only depends on the onset of a holographic BKT quantum phase transition at the boundary of a quantum critical phase, combined with the presence of irrelevant operators that couple the IR critical operator to the current. The mechanism is independent of the details of the theory undergoing the transition and also of the UV completion of the critical theory. Beyond the tuning to the quantum critical point, no additional specification of dimensions of operators or dynamical critical exponents is necessary. In the remainder of this paper we describe a specific holographic realization of the scenario that we have just outlined.

\section{Lattices and finite wavevector instabilities}
\label{sec:lattice}

A concrete holographic model realizing the mechanism outlined above can be achieved by combining three interesting ingredients that have been the focus of recent discussion:

\begin{itemize}

\item Vectorial instabilities occurring at finite wavevector $k_\star > 0$ \cite{Nakamura:2009tf, Donos:2011bh, Donos:2011qt}.

\item A (semi-) locally quantum critical sector in which all momenta are critical
\cite{Sachdev:2010um, McGreevy:2010zz, Iqbal:2011in}.

\item A lattice that is irrelevant in the IR scaling regime \cite{Hartnoll:2012rj, Horowitz:2012ky, Liu:2012tr}.

\end{itemize}

\noindent The essential idea is that the irrelevant coupling to the lattice will communicate the finite wavevector instability to the $k=0$ electrical current. The instability at $k_\star$ can have critical scaling because of the local quantum criticality.

The most important effect of a lattice is to broaden the delta function in the conductivity \cite{Hartnoll:2012rj} into a Drude peak. Impurities will also achieve this effect \cite{Hartnoll:2008hs}. Nonetheless, many of the interesting materials are believed to be very clean, and indeed, the Fermi liquid $T^2$ resistivity observed over large ranges of temperatures in these materials, away from the critical regions, presumably originates from umklapp scattering off the lattice rather than impurities plus interactions \cite{maslov}. As we have stressed repeatedly, however, in this work we are not interested in the Drude peak contribution to the conductivity. The role of the lattice for us will be to mix modes with different wavevectors.

We will consider the following 3+1 dimensional bulk model \cite{Donos:2011bh}
\begin{align}\label{lagra1}
\mathcal{L}=&\half R\,\ast 1-\half \,\ast d\varphi\wedge d\varphi-V\left(\varphi\right) \ast 1- \half \,\tau\left(\varphi\right) F\wedge\ast F-\half \,\vartheta\left(\varphi\right) F\wedge F\, ,
\end{align}
where $F=dA$. This Lagrangian describes Einstein-Maxwell theory coupled to a pseudoscalar field with both dilatonic and axionic couplings to the field strength. The corresponding equations of motion are given by
\begin{align}\label{eomi}
&R_{\mu\nu}=\partial_\mu \varphi\partial_\nu \varphi+g_{\mu\nu}\,V-\tau\,\left(\frac{1}{4}g_{\mu\nu}\,F_{\lambda\rho}F^{\lambda\rho} -F_{\mu\rho}F_{\nu}{}^{\rho}\right) \,, \nn
&d\left(\tau\ast F+\vartheta F\right)=0 \,, \nn
&d\ast d\varphi+V'*1+\frac{1}{2}\tau'\,F\wedge\ast F+\frac{1}{2}\vartheta'\,F\wedge F=0\, .
\end{align}
We will assume that the three functions $V$, $\tau$ and $\vartheta$ have the following expansions
\begin{align}\label{expans}
V=-6+\frac{1}{2}m_{s}^{2}\,\varphi^{2}+\dots,\qquad
\tau=1-\frac{n}{12}\,\varphi^{2}+\dots,\qquad
\vartheta=\frac{c_{1}}{2\sqrt{3}}\,\varphi+\dots\, .
\end{align}
It is sufficient to know the action for the pseudoscalar to quadratic order, as it will only appear as a perturbation of the background. In (\ref{expans}) we have furthermore set the asymptotic $AdS_4$ radius to $L^2 = 1/2$. In the Lagrangian (\ref{lagra1}) we already set both the Newton and Maxwell constants to unity.
All of these quantities can be scaled out of the equations of motion in this model.

The key feature of the theory (\ref{lagra1}), to be described immediately below, is that it can develop vectorial instabilities, which involve the electric current. Closely related to this fact is that the instabilities of interest occur at nonzero wavevector $k_\star$, avoiding the spontaneous generation of a homogeneous current, as we would expect. By subsequently introducing a lattice, that becomes irrelevant in the IR and does not disrupt the critical phase, we will communicate the effects of this instability to the spectral weight of the current operator, leading to strong temperature and frequency dependence in the conductivity.

\subsection{IR spectrum of excitations and instability}
\label{sec:irspectrum}

The equations of motion (\ref{eomi}) admit the following $AdS_{2}\times \mathbb{R}^{2}$ black hole solution
\be\label{eq:AdS2_bh}
ds_{4}^{2} = -f\,dt^{2}+\frac{1}{12} \left(\frac{dr^{2}}{f}+dx^{2}+dy^{2}\right) \,, \qquad A=\left(r-r_{+}\right)\,dt \,, \qquad
\varphi=0 \,.
\ee
Here the emblackening factor
\be
f=r^{2}-r_{+}^{2} \,.
\ee
The horizon of the black hole \eqref{eq:AdS2_bh} is at $r=r_{+}$ and the temperature is $T = \sqrt{3}\,r_{+}/\pi$.
The boundary of $AdS_2$ is at $r \to \infty$. This will serve as our near horizon scaling geometry,
 as per our discussion in \S \ref{sec:bkt} above, which we have further
placed at a finite temperature. That the effects of the temperature are restricted to the IR scaling geometry implies that $T \ll \mu$, the UV scale. The scaling regime described by $AdS_{2}\times \mathbb{R}^{2}$ has dynamical critical exponent $z=\infty$ and an associated ground state entropy density. This ground state entropy has tainted the reputation of $AdS_{2}\times \mathbb{R}^{2}$ as a ubiquitous tool in the applied holography effort. (Semi) local quantum criticality with $z=\infty$ is in fact compatible with a vanishing ground state entropy density if hyperscaling is violated \cite{Hartnoll:2012wm}. Holographic scaling geometries with $z=\infty$ are the only currently known holographic duals (away from the probe limit) that share at leading bulk classical order the basic property of Fermi liquids of having spectral weight at low energies but finite momentum. It seems conceivable that $AdS_{2}\times \mathbb{R}^{2}$ may have the last laugh.

As usual, perturbations about the background can be decomposed into transverse and longitudinal sectors that decouple from each other. The finite wavevector instability occurs in the transverse channel. We can write the transverse perturbations around the exact solution \eqref{eq:AdS2_bh} as
\begin{align}\label{fluctuations}
\delta g_{ty}=h_{ty}\left(t,r\right)\sin\left(kx\right) \,, \qquad \delta g_{xy}=h_{xy}\left(t,r\right)\cos\left(kx\right) \,, \nn
\delta A_{y}=a\left(t,r\right)\sin\left(kx\right) \,, \qquad \qquad \delta \varphi= w\left(t,r\right)\cos\left(kx\right) \,.
\end{align}
The equations of motion \eqref{eomi} then yield the system of linear coupled equations
\begin{align}
-k\,\partial_{t}h_{xy}+k^{2}\,h_{ty}-f \left(2\,\partial_{r}a+\partial_{r}^{2}h_{ty}\right)=0 \,, \nn
2\,\partial_{t}a+12kf\,\partial_{r}h_{xy}+\partial_{t}\partial_{r}h_{ty}=0 \,, \nn
-f^{-1}\,\partial_{t}^{2}h_{xy}+12\,\partial_{r}\left(f\partial_{r}h_{xy} \right)+kf^{-1}\,\partial_{t}h_{ty}=0 \,, \nn
-f^{-1}\,\partial_{t}^{2}a+12\,\partial_{r}\left(f\partial_{r}a \right)-12k^{2}\,a+c_{1}k\,w+12\,\partial_{r}h_{ty}=0 \,, \nn
-f^{-1}\,\partial_{t}^{2}w+12\,\partial_{r}\left(f\partial_{r}w \right)-\left(12k^{2}+m_{s}^{2}+n\right)\,w+12c_{1}k\,a=0 \,.
\end{align}
Introducing the new field $\phi_{xy}$ through
\begin{equation}
\partial_{t}\phi_{xy}=f\,\partial_{r}h_{xy} \,,
\end{equation}
the system of equations \eqref{fluctuations} leads to the linear system of equations
\begin{equation}\label{eq:system}
\left(\Box_{2}-M^{2}\right)\mathbf{v}=0 \,,
\end{equation}
with ${\bf v}=(\phi_{xy},a,w)$ and the mass matrix
\begin{equation}\label{mmfirstmod}
M^{2}=\left(\begin{array}{ccc}12\, k^{2} & 2\,k & 0 \\144\,k & 24+12\,k^{2}  & -c_{1}k \\0 & -12\,c_{1}k & 12\,k^{2}+m_{s}^{2} + n\end{array}\right)\, .
\end{equation}
Note that the Maxwell field fluctuation decouples from the remainder at zero wavenumber, $k=0$.
The Laplacian appearing in equation \eqref{eq:system} is with respect to the two dimensional metric
\begin{equation}
ds_{2}^{2}=-f\,dt^{2}+\frac{dr^{2}}{12\,f} \,.
\end{equation}

The linearly coupled system of equations \eqref{eq:system} can be diagonalized to yield three independent modes
\bea
\lefteqn{\left(\Box_{2}-\mu_{i}^{2}\right)\,g_{i}=0}\nn
&& \Rightarrow \qquad  \partial_{r}^{2}g_{i}+\frac{f^{\prime}}{f}\,\partial_{r}g_{i}-\left(\frac{1}{12\,f^{2}}\partial_{t}^{2}+\frac{1}{12\,f}\mu_{i}^{2} \right)g_{i}=0 \,,
\eea
with $\mu_{i}^{2}$ being the three eigenvalues of the mass matrix \eqref{mmfirstmod}. If we go to frequency space by writing $g_{i}(t,r)=e^{-\imath\omega t} u_{i}(r)$, the solution with infalling boundary conditions at the horizon \cite{Son:2002sd, Hartnoll:2009sz} is then
\be
u_{i}(r)=\left(\frac{2}{r_{+}}\right)^{\nu_{i}}\,\Gamma\left(a_{i} \right)\Gamma\left(1+\nu_{i}\right)f(r){}^{-\imath \frac{\omega}{4\sqrt{3}r_{+}}}r^{-a_{i}}\,{}_{2}F_{1}\left(\frac{a_{i}}{2},\frac{a_{i}+1}{2},1-\nu_{i};\frac{r_{+}^{2}}{r^{2}} \right)-\left(\nu_{i}\leftrightarrow -\nu_{i}\right) \,,
\ee
where $a_{i}=\frac{1}{2}-\imath \frac{\omega}{2\sqrt{3}r_{+}}-\nu_{i}$ and $\nu_{i}=\frac{1}{2}\sqrt{1+\frac{1}{3}\mu_{i}^{2}}$.
By expanding the above solution near the boundary of the $AdS_{2}$, i.e. as $r \to \infty$, we can read off the $AdS_{2}$ retarded Green's functions in the usual way (e.g. \cite{Faulkner:2011tm}) to obtain\footnote{The case $\nu_i = 0$, which will be of particular interest below, should properly be treated independently, with the solution to the wave equation expressed in terms of a Legendre function. Correct result are obtained by continuation of the general results to $\nu_i \to 0$. In particular, the spectra density $\text{Im} G^R \sim \w/T$ at $\nu_i = 0$.}
\begin{equation}
\mathcal{G}^R_{i}\left(\omega, T \right)=-\left(\frac{\pi}{2\sqrt{3}} T \right)^{2\nu_{i}}\frac{\Gamma\left(1-\nu_{i}\right)\Gamma\left(\frac{1}{2}-\imath\frac{\omega}{2\pi T}+\nu_{i}\right)}{\Gamma\left(1+\nu_{i}\right)\Gamma\left(\frac{1}{2}-\imath\frac{\omega}{2\pi T}-\nu_{i}\right)} \,. \label{eq:gads2}
\end{equation}
These have the expected form that is determined by the $SL(2,\R)$ symmetry of $AdS_{2}$. Again, see for instance \cite{Faulkner:2011tm}. These are the IR Green's functions that will appear in the matching formula (\ref{eq:imgsum}).
The perturbation includes the transverse current mode $\d A_y(x)$, at nonzero momentum $k$. Below we will couple this mode to the homogeneous current using a lattice.

For small $\omega$ we have
\bea
\lefteqn{\mathcal{G}^R_{i}\left(\omega\right)=-\left(\frac{\pi}{2\sqrt{3}} T \right)^{2\nu_{i}}\frac{\Gamma\left(1-\nu_{i}\right)\Gamma\left(\frac{1}{2}+\nu_{i}\right)}{\Gamma\left(1+\nu_{i}\right)\Gamma\left(\frac{1}{2}-\nu_{i}\right)}}\nn
&& +\imath\omega \frac{1}{2}\,\left(\frac{\pi}{2\sqrt{3}} \right)^{2\nu_{i}}\,T^{2\nu_{i}-1}\frac{\Gamma\left(\frac{1}{2}+\nu_{i} \right)\Gamma\left(1-\nu_{i}\right)}{\Gamma\left(\frac{1}{2}-\nu_{i}\right)\Gamma\left(1+\nu_{i}\right)}\,\tan\pi\nu_{i} \,.
\eea
If $\nu_i$ is real, then taking the imaginary part we recover the scaling with temperature that we anticipated in (\ref{eq:im}) above, using the expression (\ref{eq:delta}) for $\Delta$ in terms of $\nu$.
In general, the eigenvalues of the matrix \eqref{mmfirstmod} are slightly complicated functions of the wavenumber $k$, but are easily found numerically.

To illustrate the features of the spectrum of our theory, consider the special case where $m_{s}^{2}+n=0$.
(We will in fact consider a different case for the numerics below.) In that case we have
\begin{align}
\nu_{1}=&\frac{1}{2}\sqrt{1+4\,k^{2}} \,, \nn
\nu_{2}=&\frac{1}{2}\sqrt{5+4\,k^{2}-\frac{\sqrt{12}}{3}\sqrt{12+\left(24+c_{1}^{2}\right)k^{2}}} \,, \nn
\nu_{3}=&\frac{1}{2}\sqrt{5+4\,k^{2}+\frac{\sqrt{12}}{3}\sqrt{12+\left(24+c_{1}^{2}\right)k^{2}}} \,.\label{eq:nus}
\end{align}
From the expressions above we can see that $\nu_{2}^{2}$ has two minima at
\be
k^{\pm}_\text{min} = \pm\frac{1}{2\sqrt{12}}\,\frac{c_{1}\sqrt{48+c_{1}^{2}}}{\sqrt{24+c_{1}^{2}}} \,,
\ee
at which
\be
\nu_{2\,\text{min}} = \frac{1}{4}\sqrt{12-\frac{c_{1}^{2}}{3}-\frac{192}{24+c_{1}^{2}}} \,.
\ee
For $c_{1}>2\sqrt{6}$ we see that there is a range of $k$ for which $\nu_{2}$ is imaginary.
These are the finite wavenumber instabilities that we shall use. For our purposes, it does not matter whether or not the range of unstable momenta extends down to $k=0$. What is important is that there are finite wavenumber instabilities involving the Maxwell field $\d A_y$.

Even when the system is stable, when $0<c_{1}<2\sqrt{6}$, we see that, as opposed to the pure Einstein-Maxwell case, the dominant mode will have $\nu_{2\, \text{min}}<1/2$. This renders the finite wavenumber operators relevant in the IR. A lattice of such operators would lead to a strong backreaction on the IR geometry \cite{Hartnoll:2012rj}. In the following subsection we will instead introduce a lattice for the charge density $J^t$, that will be seen to be irrelevant in the IR.

Within the IR geometry, neither $c_1$ nor the unstable range of $k$ is tunable. The wavenumber $k$ is dimensionless in the IR because of the local criticality. However, once the $AdS_2 \times \R^2$ geometry is realized as the IR of a full asymptotically $AdS_4$ spacetime, then we shall see that $k$ must appear in the UV-dimensionless combination $k/\sqrt{J^t}$, with $J^t$ being the total charge density in the system. By varying the charge density, which can be thought of as a proxy for varying the doping of an experimental system, we effectively vary the momentum that appears in the scaling IR Green's functions. In the following subsection we shall introduce a lattice wavenumber $k_L$. By keeping the lattice fixed and varying the charge density $J^t$, we will induce a BKT quantum phase transition in modes that couple via scattering off the lattice to the homogeneous current.

The scenario outlined in the previous paragraph, and to be fleshed out below, does not quite give a realization of the phase diagram of figure \ref{fig:phased}: the system is either stable or unstable over some range of momentum and this fact cannot be tuned. What we will vary by varying $k/\sqrt{J^t}$ is whether the instability feeds through to the homogeneous current, our observable of interest. To truly realize the desired phase diagram of figure \ref{fig:phased} we need to enlarge the model to allow e.g. the parameter in the action $c_1$ or the radius of the IR $AdS_2$ metric to be tunable from the UV. We do not see an obstacle to doing this, but will continue with our slightly simpler model in this paper.

\subsection{Lattice}

The previous subsection studied perturbations of a finite temperature scaling solution to the theory (\ref{lagra1}). Following \cite{Donos:2011bh} we exhibited an instability of this background, for $c_1$ sufficiently large, over a range of momenta, $k_A < k < k_B$. In this range the exponent $\nu_2(k)$ in (\ref{eq:nus}) is imaginary. At the boundary of this range of momenta $\nu_2(k_{A/B}) = 0$, leading to the universal spectral function (\ref{eq:critical}). In this section we will couple this critical mode to the $k=0$ current by introducing a lattice that is irrelevant in the IR. We will furthermore explain how $k$ can be tuned to the critical values by varying the charge doping of the system.

The simplest and perhaps most natural lattice to consider is an `ionic lattice' in which the operator that couples to the lattice is the charge density $J^t$. That is, we introduce a spatial modulation to the chemical potential
\be\label{eq:AV}
A_t^{(0)} = \mu + V(x,y) \,.
\ee
Holographically this is implemented by the usual UV boundary condition $\lim_{r \to \infty} A_t = A_t^{(0)}$. In a purely electric background, we can consistently set the pseudoscalar $\varphi = 0$ in the theory (\ref{lagra1}). Therefore the background generated by the lattice is found by solving the Einstein-Maxwell equations (sans pseudoscalar) subject to the UV boundary condition. We shall do this numerically in a perturbative approximation in the following section. However, the essential physical features can be determined without an explicit background solution, as we proceed to show.

The charge density at any nonzero wavevector is an irrelevant perturbation about the $AdS_2 \times \R^2$ solution (\ref{eq:AdS2_bh}) of Einstein-Maxwell theory \cite{Edalati:2010pn, Hartnoll:2012rj}. Therefore, however strong the UV potential in (\ref{eq:AV}), the effects of the lattice will always be small in the far IR. The IR Green's functions computed in the previous section are thus unchanged. The role of the lattice is only to mix the unstable finite wavenumber modes with the homogeneous current mode, according to our discussion in \S \ref{sec:mechanism}. To isolate this effect from Drude momentum-relaxation physics, we take the current to propagate in the $y$ direction, but the lattice to oscillate only in the $x$ direction, with wavenumber $k_L$
\be\label{eq:periodic}
V(x,y) = V(x) = V\left(x + \frac{2\pi}{k_L}\right) \,.
\ee

According to (\ref{eq:sigma}), to obtain the conductivity we must calculate $\text{Im} \, G^R_{J^y J^y}(\w,T)$
at $k=0$. We are now taking the current to run in the $y$ rather than $x$ direction. This Green's function is holographically related to perturbations $\delta A_y$ in the bulk. Even without a lattice, in a nonzero charge density spacetime, $\delta A_y$ couples to $\d g_{ty}$. With the lattice, inspection of the equations reveals that in the full spacetime the four modes $\{\delta A_y, \delta g_{ty}, \delta g_{xy}, \delta \varphi\}$ are coupled at wavenumbers that are integer multiplies of $k_L$. These can be collected into three gauge-invariant modes for each wavenumber
\be
k_n = n \, k_L \,,
\ee
satisfying second order equations of motion. In the IR, the different wavenumber modes decouple and obey the equations that we have solved in the previous subsetion. The upshot is that the general matching formula (\ref{eq:imgsum}) becomes
\be
\text{Im} \, G^R_{J^y J^y}(\w,T) = \sum_{i,n} d^i_n \, \text{Im} \, {\mathcal G}^R_i(\w,T, k_n) \,.
\ee
Here the IR Green's functions ${\mathcal G}^R_i$ are those obtained in (\ref{eq:gads2}) above. Their wavenumber dependence is through $\nu_i(k_n)$ in (\ref{eq:nus}). The mixing is illustrated in figure \ref{fig:lattice}  below.

\begin{figure}[h]
\begin{center}
\includegraphics[height=170pt]{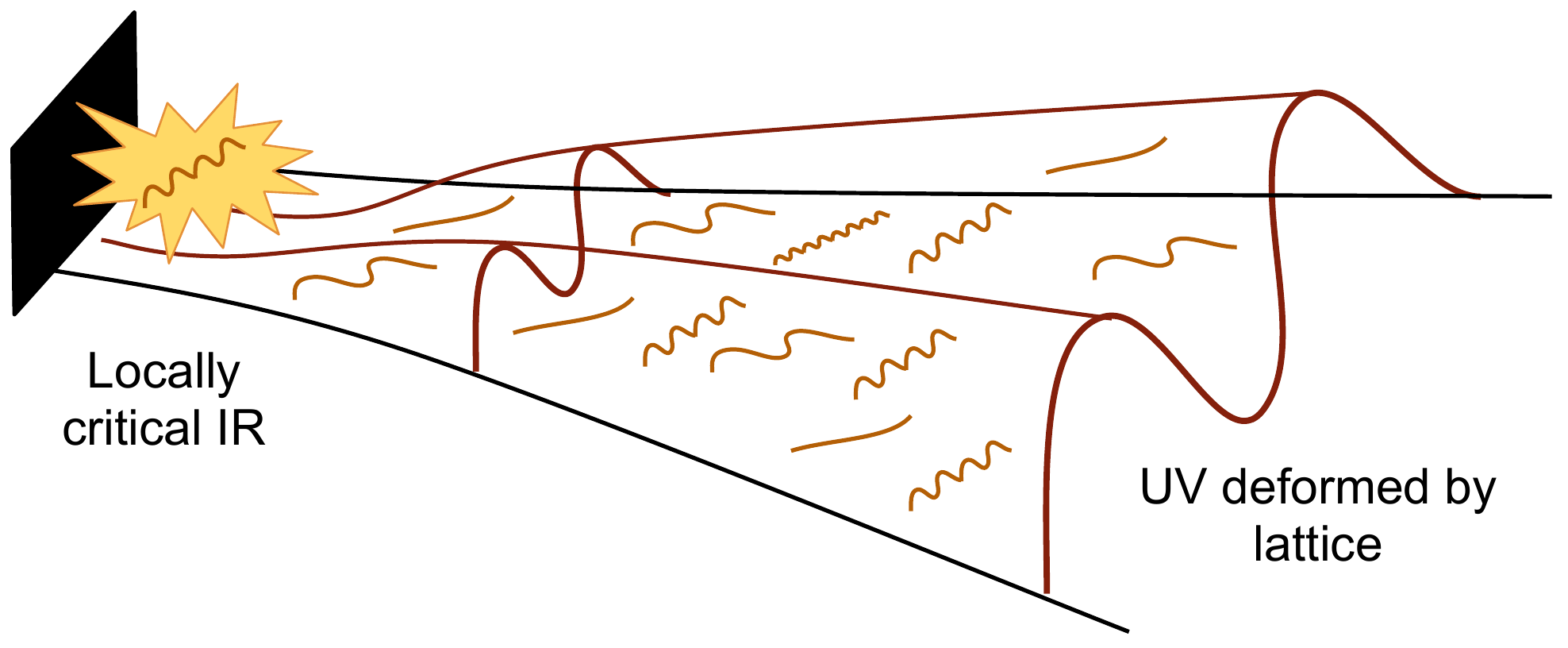}\caption{The IR is tuned to the boundary of an instability condensing finite wavenumber vectorial modes. The lattice is imposed in the UV but irrelevant in the IR. Away from the locally critical IR region, the lattice mixes modes of different wavenumber and couples the unstable mode to the homogeneous electric current.
\label{fig:lattice}}
\end{center}
\end{figure}

To tune the system across the instability identified in the previous subsection, insofar as the conductivity is concerned (see the discussion at the end of \S \ref{sec:irspectrum}), we now need to vary $k_L$ so that one of the $k_n$ becomes equal to the wavenumbers $k_A$ or $k_B$ bounding the unstable region. Physically it is presumably not feasible to tune $k_L$ itself, as the material is fixed. However, in the formulae for $\nu_i(k_n)$ in (\ref{eq:nus}), a dimensionful quantity has been suppressed and lengths have been rescaled relative to natural lengths of the UV geometry \cite{Hartnoll:2011fn}. This is possible because of the emergent $z = \infty$ scaling, under which lengths are dimensionless.
The invariant dimensionless quantity to consider, from a UV perspective, is the ratio of momentum to the square root of the charge density: $k/\sqrt{J^t}$ \cite{Hartnoll:2012rj}. The charge density $J^t$ corresponds to the total electric charge available for conduction. This can be tuned experimentally by doping the system. Doping is indeed the mechanism of tuning in the cuprate and pnictide unconventional superconductors. Thus, with $k_F$ fixed, doping the system will allow the dimensionless IR wavenumbers $k_n$ to be tuned through their critical values, communicating the BKT transition of interest into a linear in temperature resistivity (\ref{eq:linearT}).

The optical conductivities in (\ref{eq:sw}) and (\ref{eq:1w}) now show that a lattice combined with local criticality and a mode on the verge of a finite wavenumber instability leads to a significant low energy spectral weight, ranging to $\sigma \sim \w^{-1}$ at the critical point itself. A strong optical conductivity going like $\sigma \sim \w^{-2/3}$ was recently obtained in an impressive numerical holographic lattice computation in \cite{Horowitz:2012ky}. The mechanism underlying these effects are quite different. The lattice scatterings captured in \cite{Horowitz:2012ky} relaxed the current, resolving the Drude peak, while ours merely mix modes of differing transverse momenta. The tail  in the optical conductivity observed in \cite{Horowitz:2012ky} is indeed a tail of the Drude peak itself, and would presumably not be there if momentum were conserved. Our tail is an additional feature due to scattering off IR critical modes. Remarkably the scaling $\sigma \sim \w^{-2/3}$ agrees with a reported measurement in BSCYCO \cite{twothirds}. As we noted above, a $\sigma \sim \w^{-1}$ scaling has been suggested for e.g.
YBCO \cite{Varma:1989zz} and LSCO \cite{lsco}, although a sophisticated scaling analysis of the data comparable to that in BSCYCO \cite{twothirds} remains to be performed. To substantiate these similarities, a connection to the underlying dynamics of the material is required. It is certainly of interest to combine these two effects of the lattice in a single model.

\section{Numerics: lattice deformed RN black hole}

The explicit realization of the mechanism for linear in temperature resistivity, and strong optical conductivities, of the previous section depended on various genericity assumptions about modes coupling to other modes. In this section we perform a numerical analysis with a lattice that will confirm all the assumptions made. To simplify the analysis, we work perturbatively (to second order) in the strength of the lattice. This allows us to use a large system of coupled ODEs rather than PDEs. The perturbative treatment is consistent and already contains all of the physics we wish to illustrate.

\subsection{The background lattice perturbation}\label{sec:lattice_sol}

In the absence of a lattice, the full asymptotically $AdS_4$ background of the theory (\ref{lagra1}) dually describing
a finite charge density, zero temperature state of matter is the extremal Reissner-Nordstr\"om black brane \cite{Hartnoll:2011fn}
\begin{align}\label{eq:RNbh}
ds^{2}_{4}&=-f\,dt^{2}+f^{-1}\,dr^{2}+r^{2}\,\left( dx^{2}+dy^{2}\right) \,, \notag\\
A&=\,\left(1-\frac{r_{+}}{r} \right)\,dt \,, \notag\\
f&=2r^{2}-\left( 2r_{+}^{2}+\frac{1}{2}\right)\frac{r_{+}}{r}+\frac{r_{+}^{2}}{2r^{2}} \,.
\end{align}
We wish to deform by introducing a chemical potential or ``ionic'' lattice. We achieve this by setting the boundary value of the Maxwell field to be (cf. (\ref{eq:periodic}) above)
\begin{equation}\label{eq:At_asympt}
\lim_{r \to \infty} A_{t}=1+\lambda\,\cos\left(k_{L}x\right) \,.
\end{equation}
In this subsection we set up the equations describing the effect of this deformation on the static background (\ref{eq:RNbh}) to order $\lambda^2$.

Choosing to work in a radial gauge where $g_{r\mu}=0$, we can see that the components of the metric that will be
affected by the lattice deformation are $\left\{g_{tt},g_{rr},g_{xx},g_{yy} \right\}$. The nontrivial functional dependence of the metric components will be on the radial coordinate $r$ and the spatial coordinate $x$. There is a residual gauge symmetry generated by the vector $\xi=\xi^{r}\,\partial_{r}+\xi^{x}\,\partial_{x}$ with $\nabla_{r}\xi_{x}+\nabla_{x}\xi_{r}$=0. These perturbations of the background remain within Einstein-Maxwell theory (that is, the pseudoscalar is not sourced). As shown in \cite{Hartnoll:2012rj}, the $\w = 0$ and finite $k$ perturbations of the near horizon $AdS_{2}\times \mathbb{R}^{2}$ geometry in Einstein-Maxwell theory are all irrelevant. This fact strongly suggests that the $AdS_{2}\times \mathbb{R}^{2}$ IR limit is not modified by the inclusion of a lattice. The magnitude of the effects of the lattice are tunable at the boundary and become small as $r \to r_+$ in the interior. We therefore conclude that as far as the background is concerned, for small parameter $\lambda$ we can consistently treat the backreaction of the metric and gauge field perturbatively at all radii.

Expanding our field components in $\lambda$ we can write
\begin{align}\label{eq:lattice_perts}
A_{t}&=1-\frac{r_{+}}{r}+\lambda\,a^{(1,1)}\,\cos\left(k_{L}x\right)+\lambda^{2}\,\left(a^{(2,0)}+a^{(2,2)}\,\cos\left(2k_{L}x\right) \right)+O\left(\lambda^{3}\right) \,, \notag\\
g_{tt}&=-f\,\left[1+\lambda\,Q^{(1,1)}_{tt}\,\cos\left(k_{L}x\right)+\lambda^{2}\,\left(Q^{(2,0)}_{tt}+Q^{(2,2)}_{tt}\,\cos\left(2k_{L}x\right) \right) \right]+O\left(\lambda^{3}\right) \,, \notag\\
g_{rr}&=f^{-1}\,\left[1+\lambda\,Q^{(1,1)}_{tt}\,\cos\left(k_{L}x\right)+\lambda^{2}\,\left(Q^{(2,0)}_{tt}+Q^{(2,2)}_{tt}\,\cos\left(2k_{L}x\right) \right) \right]+O\left(\lambda^{3}\right) \,, \notag\\
g_{ii}&=r^{2}\,\left[1+\lambda\,Q^{(1,1)}_{ii}\,\cos\left(k_{L}x\right)+\lambda^{2}\,\left(Q^{(2,0)}_{ii}+Q^{(2,2)}_{ii}\,\cos\left(2k_{L}x\right) \right) \right]+O\left(\lambda^{3}\right) \,,
\end{align}
where $f$ is as in \eqref{eq:RNbh} and the $a$ and $Q$ functions depend on $r$ only. We have used the residual gauge freedom mentioned above to fix the functions appearing in $g_{rr}$ in terms of the functions appearing in $g_{tt}$. From
the expansion above we have, computing the surface gravity, that the temperature is still given by\footnote{We will later impose that the metric perturbations fall off at infinity, so that the normalization of time remains the same as for the $\lambda=0$ case.} $4 \pi T= \left. f^{\prime}\right|_{r=r_{+}}$. Plugging the ansatz for the perturbations \eqref{eq:lattice_perts} in the equations of motion we obtain
\begin{itemize}
\item At order $\lambda$: second order differential equations for the perturbations $\left\{Q_{ii}^{(1,1)},a^{(1,1)}\right\}$ and an algebraic equation for $Q^{(1,1)}_{tt}$.
\item At order $\lambda^{2}$: second order equations for $\left\{Q_{ii}^{(2,0)},Q_{ii}^{(2,2)},a^{(2,0)},a^{(2,2)}\right\}$, a first order equation for $Q^{(2,0)}_{tt}$ and an algebraic equation for $Q_{tt}^{(2,2)}$.
\end{itemize}
We will solve these equations numerically below. At this early point we note that for our purposes we will not need to solve for the functions $Q_{aa}^{(2,2)}$, where $a = t,x,y$, and $a^{(2,2)}$ but it should be noted that they are non-trivial by consistency of the equations of motion. In order to fully define the boundary value problem we now need to impose boundary conditions which will give us a regular near horizon geometry and with the right sources at the boundary of $AdS_{4}$.

The horizon is a singular point of the equations of motion and regularity of the solution imposes the following near horizon expansions at $r=r_{+}$
\begin{align}\label{eq:nhb_exp}
Q_{aa}^{(i,j)}&=Q_{aa(0)}^{(i,j)}+Q_{aa(1)}^{(i,j)}\,\left(r-r_{+}\right)+\cdots,\quad a=t,x,y \,, \notag\\
a^{(i,j)}&=a^{(i,j)}_{(0)}\,\left(r-r_{+}\right)+a^{(i,j)}_{(1)}\,\left(r-r_{+}\right)^{2}+\cdots \,.
\end{align}
Plugging the near horizon expansion \eqref{eq:nhb_exp} into the equations of motion, we discover that the constants $a^{(i,j)}_{(0)}$,  $Q_{tt(0)}^{(2,0)}$ and $Q_{ii(0)}^{(i,j)}$ are constants of integration as well as a combination of $Q_{ii(1)}^{(1,1)}$ which we take to be the sum $s^{(1,1)} \equiv Q_{xx(1)}^{(1,1)}+Q_{yy(1)}^{(1,1)}$.

On the other hand, at infinity we have the following schematic expansions
\begin{align}\label{eq:asb_exp}
a^{(i,j)}&=A_{(0)}^{(i,j)}+A_{(1)}^{(i,j)}\,r^{-1}+\cdots \,, \notag\\
Q_{xx}^{(1,1)}&=G_{xx(-1)}^{(1,1)}\,r+G_{xx(0)}^{(1,1)}+\cdots+G_{xx(3)}^{(1,1)}\,r^{-3}+\cdots \,, \notag\\
Q_{yy}^{(1,1)}&=G_{xx(-1)}^{(1,1)}\,r+G_{yy(0)}^{(1,1)}+\cdots+5\,G_{xx(3)}^{(1,1)}\,r^{-3}+\cdots \,, \notag\\
Q_{ii}^{(2,0)}&=G_{ii(0)}^{(2,0)}+\cdots+G_{ii(3)}^{(2,0)}\,r^{-3}+\cdots \,, \notag\\
Q_{tt}^{(2,0)}&=G_{tt(0)}^{(2,0)}+\cdots \,.
\end{align}
In order to source the gauge field corresponding to the ionic lattice deformation \eqref{eq:At_asympt} we need to set $A_{(0)}^{(i,j)}=\delta^{i1}\delta^{j1}\,\lambda$, which uniquely fixes the sources of the gauge field. The ionic lattice is the only deformation of the Reissner-Nordstrom black hole we would like to turn on and therefore we should also set $G_{xx(-1)}^{(1,1)}=G_{ii(0)}^{(1,1)}=G_{ii(0)}^{(2,0)}=G_{tt(0)}^{(2,0)}=0$.

To conclude this section we note that we will have a total of six second order differential equations\footnote{As mentioned before, we don't need to solve for the functions $Q^{(2,2)}_{aa}$ and $a^{(2,2)}$.} in $\left\{Q_{ii}^{(1,1)},Q_{ii}^{(2,0)},a^{(1,1)},a^{(2,0)}\right\}$ and one first order differential equation in $Q_{tt}^{(2,0)}$. We have eight constants of integration from the horizon $\left\{Q_{ii(0)}^{(1,1)},s^{(1,1)},Q_{aa(0)}^{(2,0)},a_{(0)}^{(1,1)},a_{(0)}^{(2,0)}\right\}$ and five constants of integration from the asymptotic data $\left\{G_{xx(3)}^{(1,1)},G_{ii(3)}^{(2,0)},A_{(1)}^{(1,1)},A_{(1)}^{(2,0)}  \right\}$. This combination of differential order and number of constants of integration gives a unique solution  for any given temperature. We will describe the numerical shooting method we are using in \S \ref{sec:numerics}.

\subsection{The $\lambda^{2}$ corrected current-current correlator}
\label{sec:corelator_cor}

Given the background whose construction we have just described, we now wish to compute the retarded Green's function $G^R_{J^y J^y}(\w,T)$ at $k=0$ in the deformed background. As will immediately become clear, for a source for $J^y$ at $k=0$ to induce a response also in $J^y$ at $k=0$ via scattering off a finite wavenumber mode, it is necessary to go to second order in the lattice strength, which is why we have also considered the background to order $\lambda^2$. To compute the Green's function, we consider the following time dependent perturbation
\begin{align}\label{eq:current_pert}
\delta A_{y}&=e^{-\imath \omega t}\left[ a_{y}^{(0,0)}+\lambda\,a_{y}^{(1,1)}\,\cos\left(k_{L}x\right)+\lambda^{2}\,\left(a_{y}^{(2,0)}+a_{y}^{(2,2)}\,\cos\left(2k_{L}x\right) \right)\right]+O\left(\lambda^{3}\right) \,, \notag\\
\delta\varphi&=e^{-\imath \omega t}\left[\lambda\,\varphi^{(1,1)}\,\sin\left(k_{L}x\right)+\lambda^{2}\,\varphi^{(2,2)}\,\sin\left(2k_{L}x \right)\right]+O\left(\lambda^{3}\right) \,, \notag\\
\delta g_{ty}&=e^{-\imath \omega t}r^{2}\left[g_{ty}^{(0,0)}+\lambda\,g_{ty}^{(1,1)}\,\cos\left(k_{L}x\right)+\lambda^{2}\,\left(g_{ty}^{(2,0)}+g_{ty}^{(2,2)}\,\cos\left(2k_{L}y \right) \right)\right]+O\left(\lambda^{3}\right) \,, \notag\\
\delta g_{xy}&=e^{-\imath \omega t}r^{2}\left[\lambda\,g_{xy}^{(1,1)}\,\sin\left(k_{L}x\right)+\lambda^{2}\,g_{xy}^{(2,2)}\,\sin\left(2k_{L}x \right)\right]+O\left(\lambda^{3}\right) \,,
\end{align}
with all the functions $\Phi^{(i,j)}$ depending on the radius $r$ only. The perturbation described by \eqref{eq:current_pert} is consistent and we chose again to work in radial gauge. We will fix the constants of integration in this gauge and then perform a large gauge transformation that will bring the perturbation into a form where it will be manifest that we are only sourcing the $k=0$ component of the current $J_{y}$.

Plugging the perturbation \eqref{eq:current_pert} into the equations of motion we obtain
\begin{itemize}
\item Second order equations for $\left\{a_{y}^{(i,j)},\varphi^{(i,j)},g_{xy}^{(i,j)} \right\}$ .\notag 
\item First order equations for $\left\{g_{ty}^{(i,j)}\right\}$ .
\end{itemize}
Regular, infalling boundary conditions at the horizon gives the near horizon expansion
\begin{align}\label{eq:nh_exp}
\varphi^{(i,j)}&=f^{-\frac{\imath \omega}{4\pi T}}\,\left[\varphi_{(0)}^{(i,j)}+\varphi_{(1)}^{(i,j)}\,\left(r-r_{+}\right)+\cdots\right] \,, \notag\\
a_{y}^{(i,j)}&=f^{-\frac{\imath \omega}{4\pi T}}\,\left[a_{y(0)}^{(i,j)}+a_{y(1)}^{(i,j)}\,\left(r-r_{+}\right)+\cdots\right] \,, \notag\\
g_{xy}^{(i,j)}&=f^{-\frac{\imath \omega}{4\pi T}}\,\left[g_{xy(0)}^{(i,j)}+g_{xy(1)}^{(i,j)}\,\left(r-r_{+}\right)+\cdots\right] \,, \notag\\
g_{ty}^{(i,j)}&=f^{-\frac{\imath \omega}{4\pi T}}\left(r-r_{+}\right)\,\left[g_{ty(0)}^{(i,j)}+g_{ty(1)}^{(i,j)}\,\left(r-r_{+}\right)+\cdots\right] \,.
\end{align}
Regularity is seen to imply that all the constants of integration are fixed in terms of $\left\{\varphi_{(0)}^{(i,j)}, a_{y(0)}^{(i,j)},g_{xy(0)}^{(i,j)}\right\}$. In particular, the $g_{ty(0)}^{(i,j)}$ are not independent.

The general asymptotic form of the solutions as $r \to \infty$ is found to be\footnote{We chose to have $m_{s}^{2}=-4$ in the action \eqref{lagra1}. Reinstating the asymptotic $AdS_4$ radius $L$, this corresponds to a mass squared $m^2 L^2 = - 2$.}
\bea\label{eq:as_exp}
\varphi^{(i,j)} & = & \displaystyle \frac{\varPhi^{(i,j)}_{(1)}}{r}+\frac{\varPhi^{(i,j)}_{(2)}}{r^{2}}+\cdots \,, \qquad g^{(i,j)}_{xy} = G^{(i,j)}_{xy(0)}+\cdots +\frac{G^{(i,j)}_{xy(3)}}{r^{3}}+\cdots \,,   \notag\\
a^{(i,j)}_{y} &  = & \displaystyle A^{(i,j)}_{y(0)}+\frac{A^{(i,j)}_{y(1)}}{r}+\cdots \,,  \qquad g^{(i,j)}_{ty}=G^{(i,j)}_{ty(0)}+\cdots \,.
\eea
For the bulk pseudoscalar $\varphi$, the absence of a source requires that we impose
$\varPhi^{(i,j)}_{(1)}=0$ for a $\Delta=2$ operator or $\varPhi^{(i,j)}_{(2)}=0$ for a $\Delta=1$ operator. We wish to source only the homogeneous mode of the $A_{y}$ component of the Maxwell field and furthermore at leading order in $\lambda$ only, giving the boundary condition $A^{(i,j)}_{y(0)}=\delta^{i,0}\delta^{j,0}$.

The situation is slightly more involved with the constants $G^{(i,j)}_{xy(0)}$ and $G^{(i,j)}_{ty(0)}$. The first thought would be that one needs to set all these constants equal to zero since they represent non-normalisable modes for the bulk fields. A simple counting, however, reveals that this would overdefine the boundary value problem. 
At the horizon we saw that $g_{ty(0)}^{(i,j)}$ was fixed by regularity in terms of the other integration constants. Consequently, at the boundary,
the constants $G^{(i,j)}_{ty(0)}$ will be uniquely fixed after specifying $G^{(i,j)}_{xy(0)}$, and demanding regularity at the horizon, and will in general be nonzero.

We will instead choose the constants $G^{(i,j)}_{ty(0)}$ and $G^{(i,j)}_{xy(0)}$ such that after a large coordinate transformation we will only have a source for the current. More specifically we impose
 \begin{equation}\label{eq:G_conds}
j\,k\,G^{(i,j)}_{ty(0)}-\imath\, \omega \, G^{(i,j)}_{xy(0)}=0 \,.
 \end{equation}
We then perform the coordinate transformation
\begin{align}\label{eq:large_diff_transf}
y\rightarrow y+e^{-\imath \omega t}\sum_{i,j} \lambda^{i} h^{(i,j)}\,\cos\left(jkx \right) \,,
\end{align}
where $h^{(i,j)}$ are functions of $r$ only. These functions are required to satisfy:
\begin{itemize}
\item Near the $AdS_{4}$ boundary they have the fall off
\begin{equation}
h^{(i,j)}\approx \frac{1}{j \, k}G_{xy(0)}^{(i,j)}+O\left(r^{-5}\ln r\right) \,.
\end{equation}
\item Near the horizon $r=r_{+}$ they have the infalling behavior
\begin{equation}
h^{(i,j)}=f^{-\frac{\imath \omega}{4\pi T}}\left(r-r_{+}\right)\,\left[h_{(0)}^{(i,j)}+h_{(1)}^{(i,j)}\,\left(r-r_{+}\right)+\cdots \right] \,.
\end{equation}
\end{itemize}
The fall off condition in the UV:
\begin{itemize}
\item Removes the non-normalisable terms from the $g_{ty}$ and $g_{xy}$ components of the metric provided we satisfy \eqref{eq:G_conds}.
\item It doesn't change the values of the VEVs for the various fields appearing in our ansatz.
\item It doesn't introduce new deformations coming from the $g_{ri}$ components.
\end{itemize}

A particularly natural example of a coordinate transformation satisfying all of the above properties is one that maps our perturbations into the
de Donder-Lorentz gauge
\begin{equation}\label{eq:DeDonder_gauge}
\nabla_{\mu}\delta g^{\mu\nu}=0 \,,\quad \nabla_{\mu}\delta A^{\mu}=0 \,.
\end{equation}
Here $\nabla$ is the connection with respect to the background solution we presented in \S \ref{sec:lattice_sol}. The gauge conditions \eqref{eq:DeDonder_gauge} become second order differential equations for the functions $h^{(i,j)}$. The resulting solution satisfies all of our above requirements.

The quantity we are interested in calculating is the optical conductivity (\ref{eq:sigma}). After taking into account the source of the gauge field specified by $A^{(i,j)}_{y(0)}=\delta^{i,0}\delta^{j,0}$, the retarded Green's function function is given by (up to a constant of proportionality)
\begin{equation}
G^R_{J^y J^y}\left(\omega,k=0\right)= \sum_{i=0}\lambda^{i}\,A_{y(1)}^{(i,0)} \,.
\end{equation}
As we will explain in more detail in \S \ref{sec:numerics}, for our purposes it will be sufficient to just consider the second order correction in $\lambda$
\begin{equation}\label{eq:curres}
G^R_{J^y J^y}\left(\omega,k=0\right)= A_{y(1)}^{(0,0)}+\lambda^{2}\,A_{y(1)}^{(2,0)} \,.
\end{equation}
The first term $A^{(0,0)}_{y(1)}$ is the, by now well understood, current-current correlator evaluated in the extremal electric Reissner-Nordstr\"om background in Einstein-Maxwell theory \cite{Hartnoll:2009sz}.

Our aim is therefore to solve for the function $a_{y}^{(2,0)}$ and more specifically we would like to extract the constant $A_{y(1)}^{(2,0)}$ from it. After staring at the equations of motion one will notice that in order to achieve that we will not need to solve for the functions $a_{y}^{(2,2)}$, $g_{ty}^{(2,2)}$, $g_{xy}^{(2,2)}$ and $\varphi^{(2,2)}$. The eight functions we will be solving for are $a_{y}^{(0,0)}$, $a_{y}^{(1,1)}$, $a_{y}^{(2,0)}$, $\varphi^{(1,1)}$, $g_{ty}^{(0,0)}$, $g_{ty}^{(1,1)}$, $g_{ty}^{(2,0)}$ and $g_{xy}^{(1,1)}$ which satisfy five second order differential equations and three first order ones. In order to find a solution we have five constants of integration from the horizon data $\left\{a_{y(0)}^{(0,0)},a_{y(0)}^{(1,1)},a_{y(0)}^{(2,0)},\varphi_{(0)}^{(1,1)},g_{xy(0)}^{(1,1)} \right\}$ and eight constants of integration from the asymptotic boundary $\left\{A_{y(1)}^{(0,0)},A_{y(1)}^{(1,1)},A_{y(1)}^{(2,0)},\varPhi_{(1)}^{(1,1)},G^{(1,1)}_{xy(3)},G^{(0,0)}_{ty(0)},G^{(1,1)}_{ty(0)},G^{(2,0)}_{ty(0)}  \right\}$. This counting reveals that we will now have a unique solution for the perturbation once the background and the source have been fixed, including the frequency $\omega$.

\subsection{Numerical results}
\label{sec:numerics}

Having set up the problem we would like to solve, we now employ a numerical shooting method appropriate for the case at hand. As we have outlined in \S \ref{sec:lattice_sol}, the first step of the calculation is to correct the background up to second order in $\lambda$. To achieve this, we solve the equations of motion perturbatively order by order in $\lambda$ which, as we explained, boils down to solving a system of six second order and one first order differential equations. The background solution will be uniquely specified by a set of thirteen constants $c_{1}=\left\{Q_{ii(0)}^{(1,1)},s^{(1,1)},Q_{aa(0)}^{(2,0)},a_{(0)}^{(1,1)},a_{(0)}^{(2,0)}\right\}$ and $c_{2}=\left\{G_{xx(3)}^{(1,1)},G_{ii(3)}^{(2,0)},A_{(1)}^{(1,1)},A_{(1)}^{(2,0)}  \right\}$.

In order to specify these constants, we numerically integrate the equations of motion from a distance close to the horizon $r=r_{+}+\epsilon$ up to a finite radius in the bulk $r=r_{m}$, using the near horizon expansion \eqref{eq:nhb_exp} as the initial conditions at $r=r_{+}+\epsilon$. Following this procedure allows us to express the functions and their derivatives at $r=r_{m}$ as a function of the constants of integration $c_{1}$. We also integrate numerically the equations from a large radius $r=R$ down to the same point $r=r_{m}$ using the asymptotic expansion \eqref{eq:asb_exp}, this time expressing the functions and their derivatives as a function of the constants of integration $c_{2}$. In order to get a solution in the full bulk we need to match the values of the functions and their derivatives we have found at $r=r_{m}$ by suitably adjusting the constants $c_{1}$ and $c_{2}$. A simple counting shows that this solution for the constants $c_{1}$ and $c_{2}$ will be unique i.e. the number of equations we will have from matching the two solutions will be thirteen.

Once we have specified the background we follow the same procedure to solve for the perturbation we described in \S \ref{sec:corelator_cor}. This time we will be using the near horizon expansion \eqref{eq:nh_exp} and the asymptotic expansion \eqref{eq:as_exp}. After matching the two solutions in the middle of the bulk, as we did for the background solution, we will have uniquely fixed the other thirteen constants of integration: the background constants $\left\{a_{y(0)}^{(0,0)},a_{y(0)}^{(1,1)},a_{y(0)}^{(2,0)},\varphi_{(0)}^{(1,1)},g_{xy(0)}^{(1,1)} \right\}$ and the fluctuation constants $\left\{A_{y(1)}^{(0,0)},A_{y(1)}^{(1,1)},A_{y(1)}^{(2,0)},\varPhi_{(1)}^{(1,1)},G^{(1,1)}_{xy(3)},G^{(0,0)}_{ty(0)},G^{(1,1)}_{ty(0)},G^{(2,0)}_{ty(0)}  \right\}$. In total we will have to specify twenty six constants of integration, thirteen for the background and thirteen for the perturbation. The constant we are ultimately interested in is $A_{y(1)}^{(2,0)}$ which will be precisely the $\lambda^{2}$ correction of the current-current two-point function, according to (\ref{eq:curres}). The $\lambda^{2}$ correction to the conductivity will then be given by
\begin{equation}\label{eq:cond_cor}
\sigma^{(2)}=\frac{A_{y(1)}^{(2,0)}}{\imath \, \omega} \,.
\end{equation}

\begin{figure}[h]
\begin{center}
\includegraphics[height=160pt]{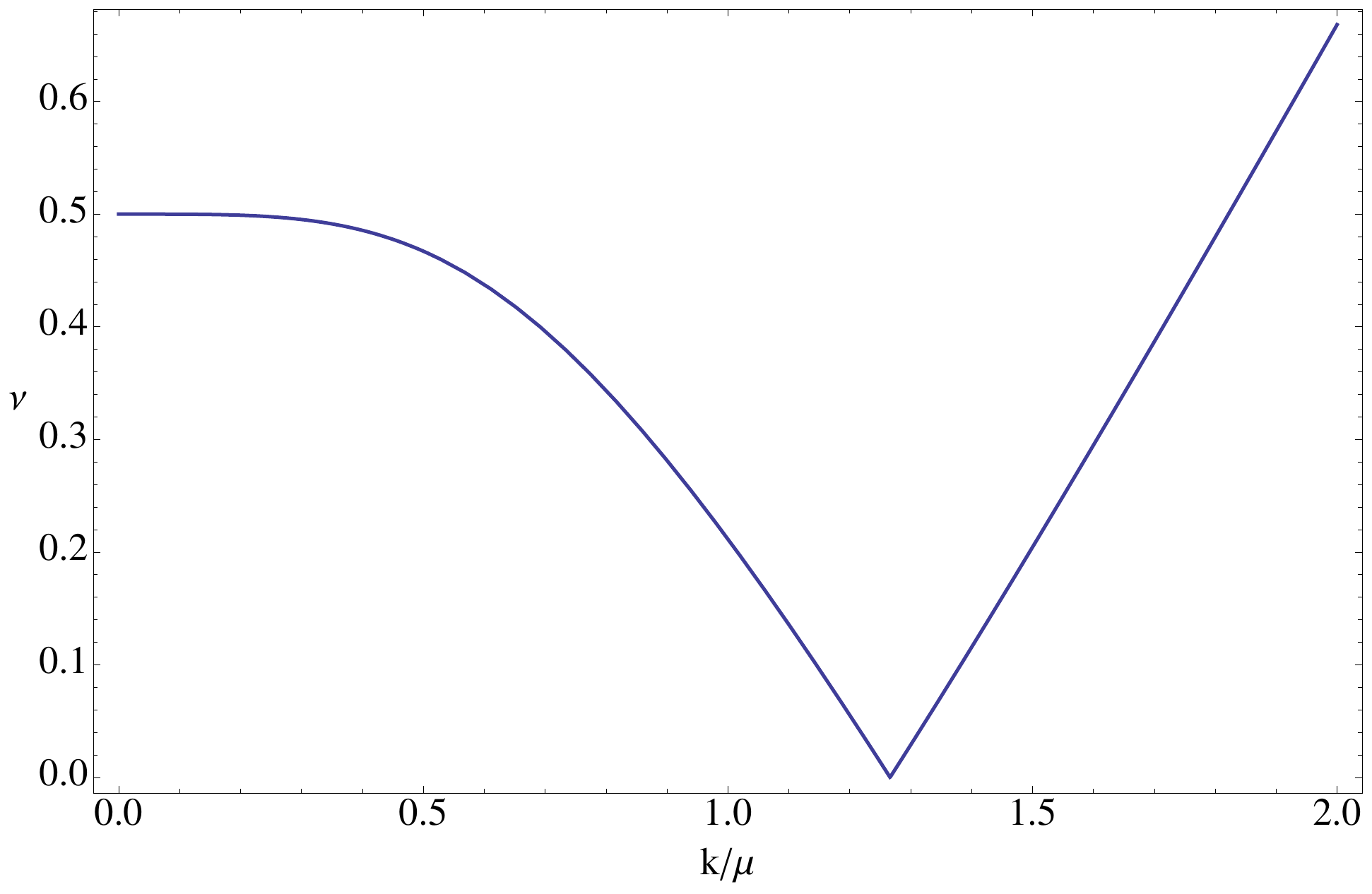}\caption{Plot of the dominant exponent $\nu$ as a function of $k$ for $n=36$, $m_s^2 = -4$ and $c_{1}\approx 8.47$ illustrating the existence of a mode with $\nu=0$ at $k_{c}\approx 1.27$, and no unstable modes. \label{fig:nu}}
\end{center}
\end{figure}

From the spectral analysis of \S \ref{sec:irspectrum} and the matching argument, we expect that if an IR relevant or unstable operator mixes in the perturbation, we should see its imprint as a strong contribution to the d.c.$\,$conductivity as we lower the system's temperature approaching extremality. To illustrate this we have fixed $n=36$ and $c_{1}\approx 8.47$ in  the Lagrangian density \eqref{lagra1}, recall we already put $m_s^2 = -4$. This choice gives exactly one value of momentum $k_{c}\approx 1.27$ for which the corresponding IR mode saturates the $AdS_{2}$ BF bound, and no unstable modes.\footnote{We have checked that for these values of the coefficients, there is also no instability localized away from the near horizon region.} The explicit expressions for the spectrum around the $AdS_{2}\times \mathbb{R}^2$ background solution is more complicated than the case with $n=4$ that we chose for illustrative purposes in section \S \ref{sec:irspectrum}. In figure \ref{fig:nu} we present the dominant $\nu(k)$ for this specific case demonstrating that at precisely $k_{c}$ we have a marginally stable mode with $\nu=0$.
This tuning to the boundary of instability is one way to address the fact, discussed at the end of \S \ref{sec:irspectrum}, that the model we are studying typically has an instability, with the effect of dialing $k_L/\sqrt{J^t}$ being to determine whether the instability manifests itself in the conductivity or not.
Though we will quote values of $k_L$ in this section, as measured in the IR theory which as $z=\infty$ and hence dimensionless momenta, we should recall from \S \ref{sec:irspectrum} that the invariant UV quantity is $k_L/\sqrt{J^t}$, that can be tuned via doping. Ultimately one would like to extend the model so that the system can be tuned through the instability.

\begin{figure}[h]
\begin{center}
\includegraphics[height=180pt]{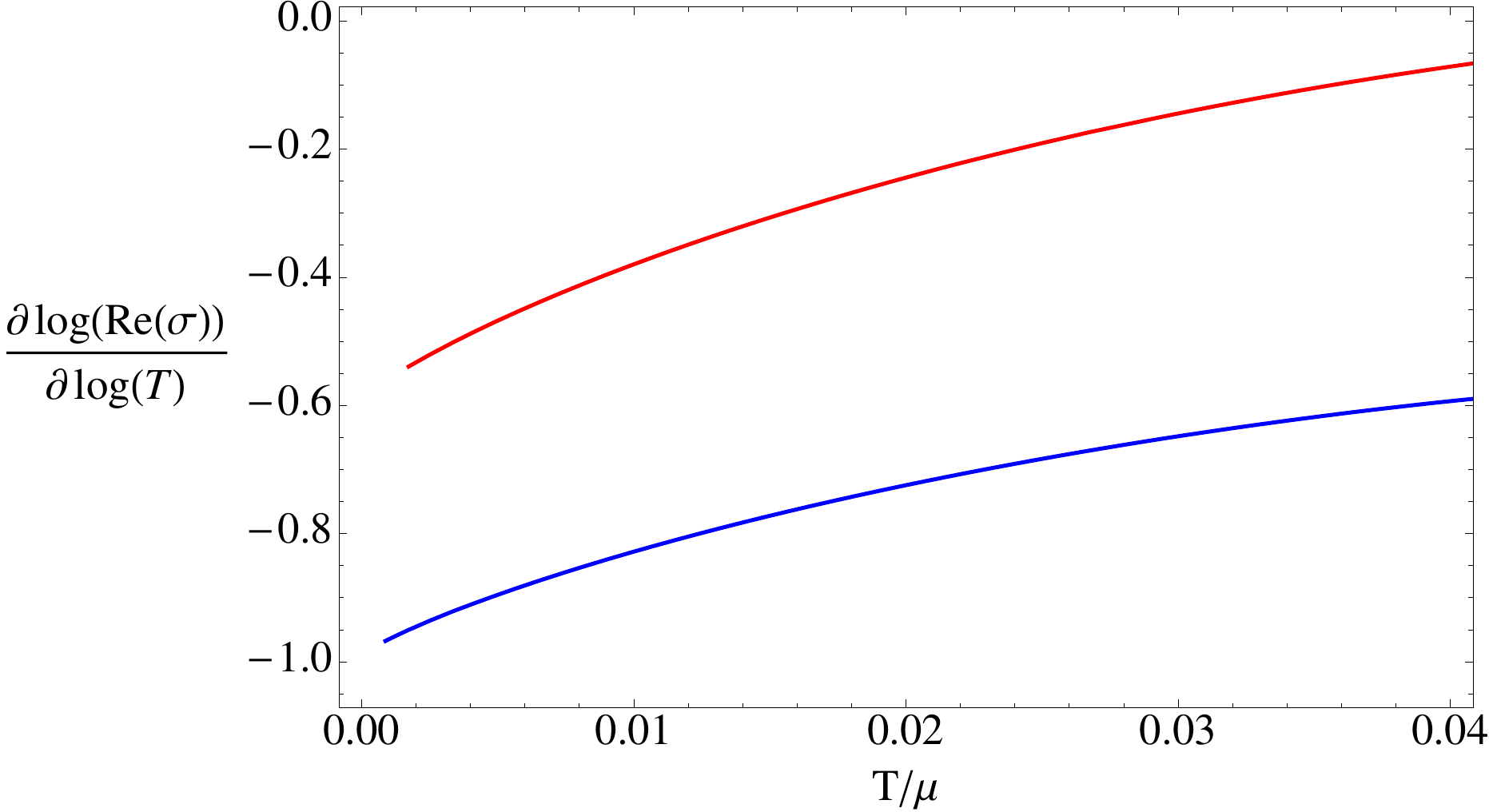}\caption{Plot of the logarithmic derivative of the temperature dependence of the d.c.$\,$conductivity for $n=36$ and $c_{1}\approx 8.47$. The top (red) curve has $k_L = 1.5$ and the bottom (blue) curve has $k_L = 1.27$.
These correspond to $\nu =0$ and $\nu \approx 0.2$, respectively. At low temperatures the expected scaling $\sigma^{(2)}\sim T^{2\nu-1}$ is recovered. \label{fig:dlogsdlogt}}
\end{center}
\end{figure}

Fixing a small value for $\omega$ (we take $\w/\mu = 10^{-6}$) and solving the equations of motion as we described above, we have calculated the correction to the d.c.$\,$conductivity due to lattice induced scatterings and we have plotted the result against temperature in figure \ref{fig:dlogsdlogt}. As we have explained in previous sections, the expectation is that at low temperatures this should be dominated by a power law $\sigma^{(2)}\sim T^{2\nu-1}$. The behaviour we observe for the two different values of lattice momentum we chose to study precisely matches our expectations.
In particular, the critical case gives $\sigma \sim 1/T$. We see that quite low temperatures are necessary in this model to reach the true scaling regime. These lower temperatures are costly to reach numerically. Figure \ref{fig:dlogsdlogt} confirms that the various genericity assumptions we have made in the matching arguments are legitimate.

\begin{figure}[h]
\begin{center}
\includegraphics[height=140pt]{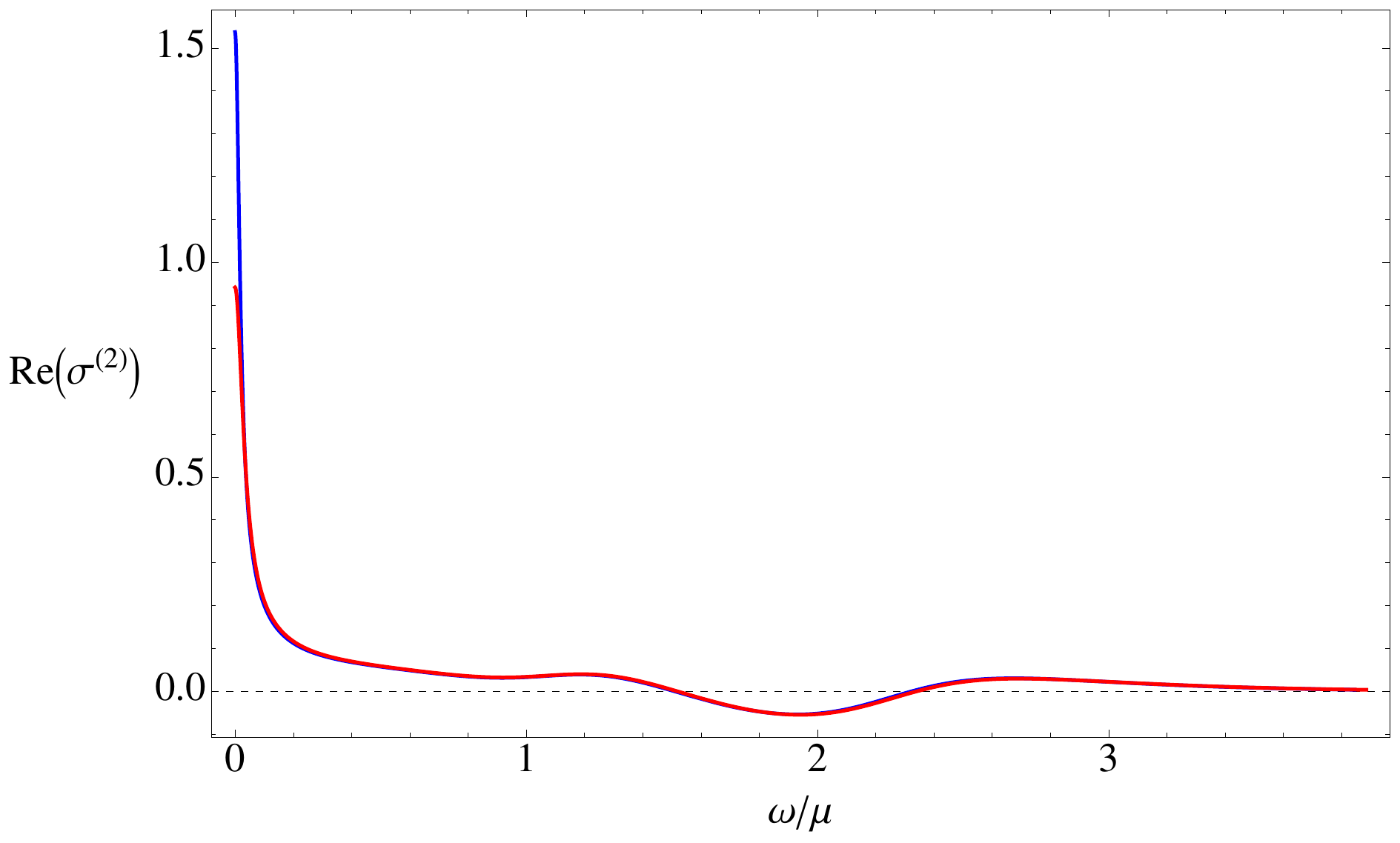}\includegraphics[height=146pt]{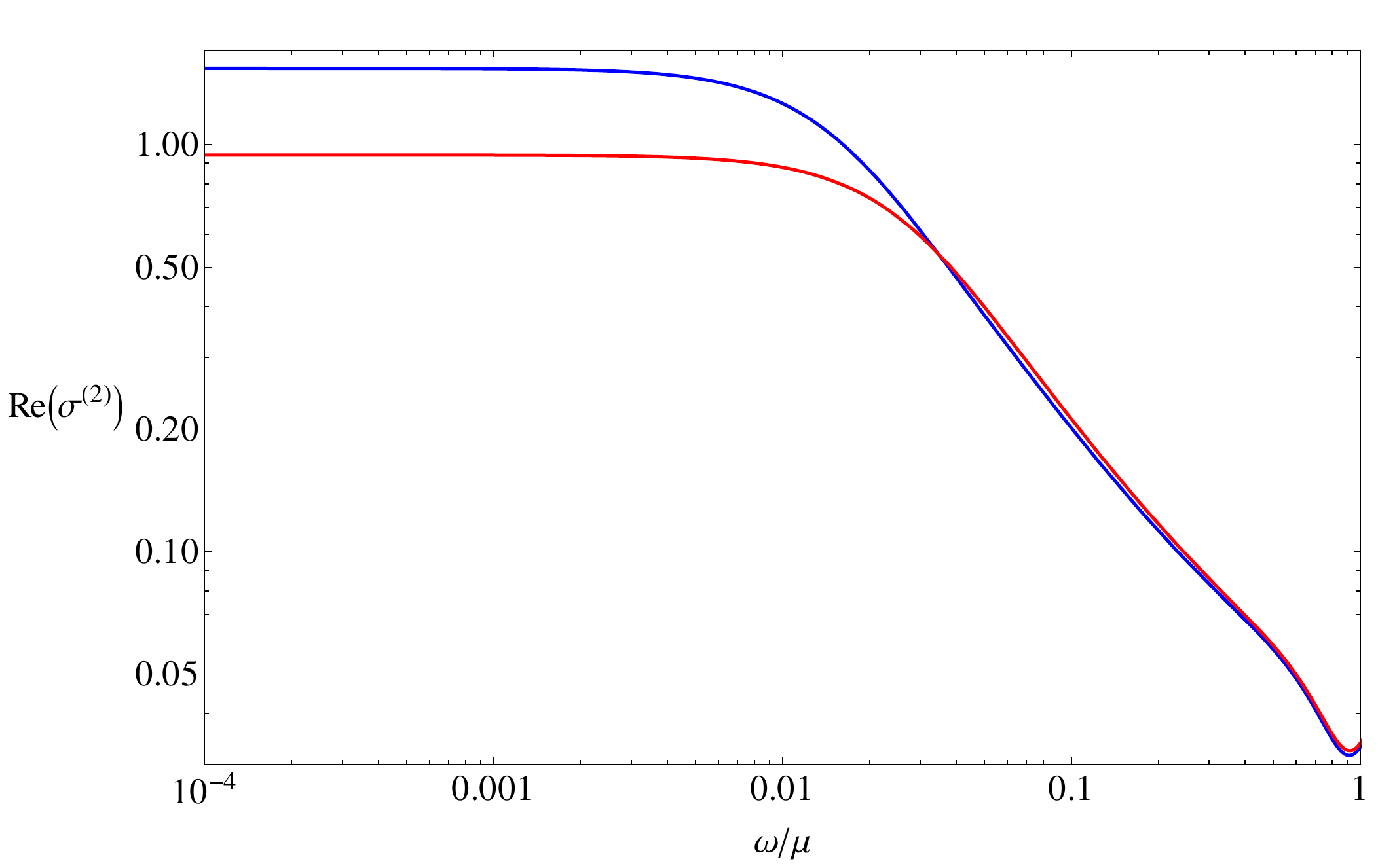}\caption{The correction to the real part of the conductivity $\mathrm{Re}\left(\sigma^{(2)} \right)$ for $n=36$, $m_s^2 = -4$ and $c_{1}\approx 8.47$ at $k_{L}=1.27$. The red (lower) curve has $T/\mu \approx 0.01$ and the blue (top) curve has $T/\mu \approx 0.006$. The left plot shows the redistribution of spectral weight at interband frequencies, while the right, log-log, plot shows the expected intermediate-low frequency scaling regime. \label{fig:OptCond}}
\end{center}
\end{figure}

We now turn to the correction to the optical conductivity. For the same values of the constants $n$ and $c_{1}$ as before, we have fixed the lattice momentum to be precisely the value for which we have the marginally stable IR mode participating in the perturbation. We have plotted the correction of the optical conductivity as a function of the frequency $\omega$ in figure \ref{fig:OptCond} for two different low values of temperature $T$. We present these results in two plots. The first shows the optical conductivity up to high frequencies. The features one notices are an extended tail at low values of the frequency $\omega$ and a regime of negative values. The negative regime implies that spectral weight is being transferred at interband energy scales. We will see in the following paragraph however that most of the spectral weight in the broad tail at intermediate energy scales is being extracted from the Drude peak. The second plot shows that the broad tail exhibits a power law falloff, again in agreement with our general expectations and matching arguments. If we fit the power law, we do not quite find the expected $\sigma \sim 1/\w$ for this critical $k_L$. This is because we are not at low enough temperatures to have reached the true scaling regime, as we saw in figure \ref{fig:dlogsdlogt}. We have checked that as the temperature is lowered, the scaling tends towards the anticipated $\sigma \sim 1/\w$.
We can recall that the zeroth order optical conductivity (without a lattice), to which the results of figure \ref{fig:OptCond} should be added, shows a featureless increase from zero at low frequencies to a constant at frequencies above $\w$ \cite{Hartnoll:2009sz}. In fact, it is a general feature of known extremal holographic backgrounds that the optical conductivity vanishes as a positive power of $\w$ at low frequencies \cite{Iizuka:2011hg}. Additional dynamics, such as the mixing with a critical mode we have implemented here, seems to be necessary to obtain an extended quantum critical tail such as that in figure \ref{fig:OptCond}. We once again emphasize that this tail is not a Drude peak, there is still a delta function at $\w = 0$, as we now discuss.

\begin{figure}[h]
\begin{center}
\includegraphics[height=140pt]{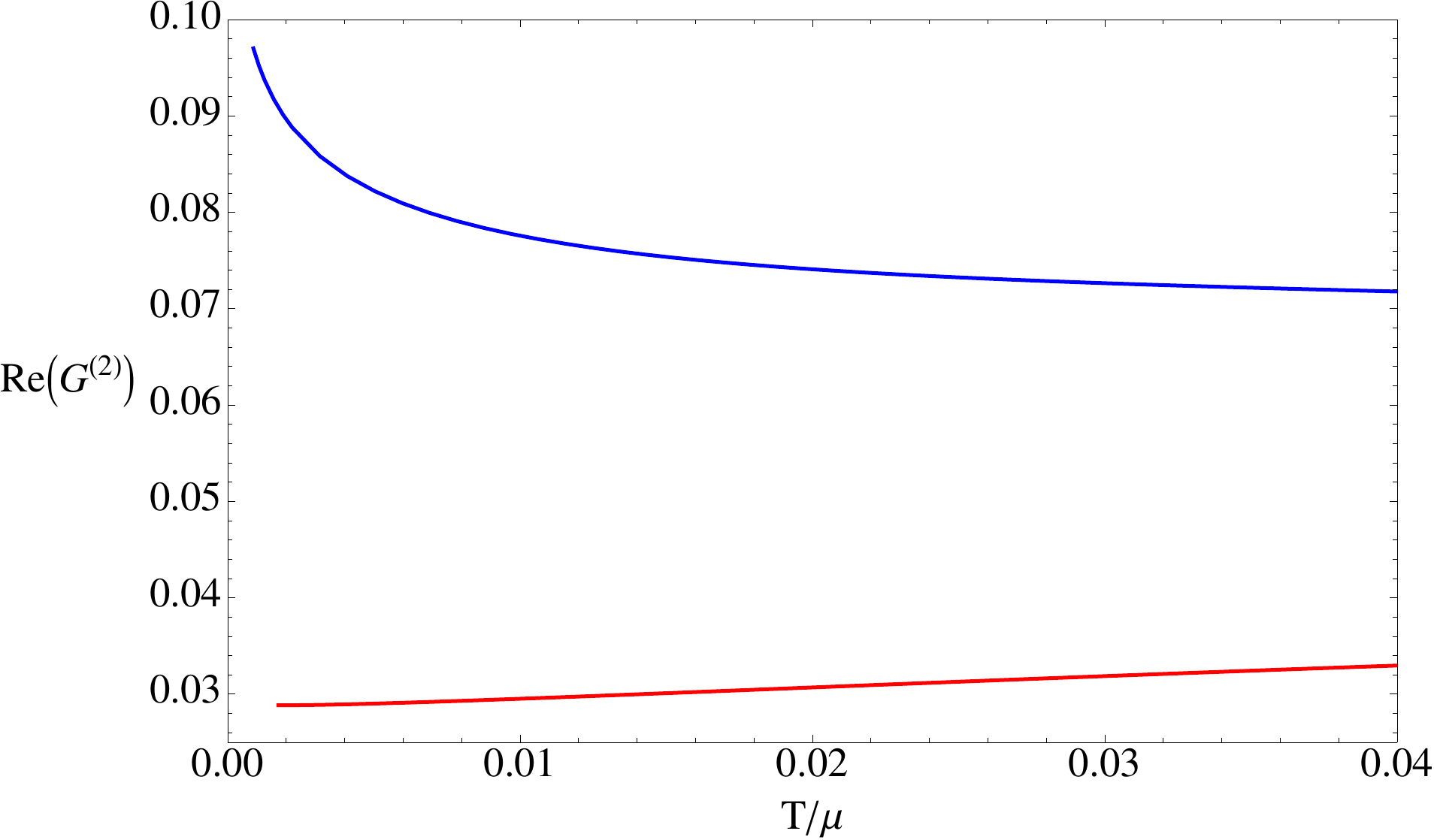}\includegraphics[height=139pt]{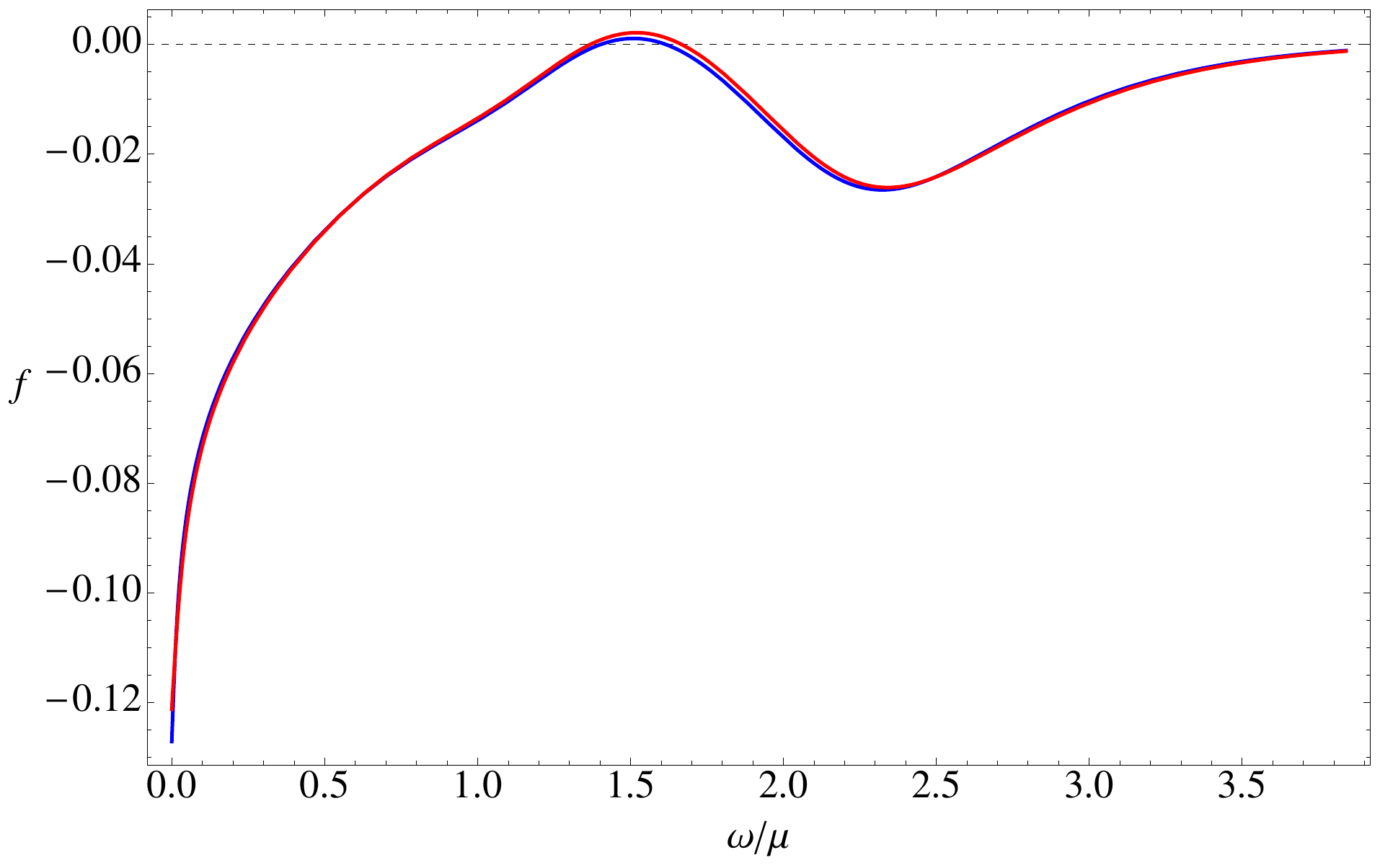}\caption{Spectral weight transfer. The plots have
$n=36$, $m_s^2 = -4$ and $c_{1}\approx 8.47$. Left, the zero frequency limit of the real part of the Green's function correction $G^{(2)}_{J^yJ^y}$ as a function of temperature. The bottom (red) curve has $k_L = 1.5$ and the top (blue) curve has $k_L = 1.27$. The negative of this curve is the spectral weight extracted from the Drude peak. Right, the integrated spectral weight defined in \eqref{eq:f_definition}. The red curve has $T/\mu \approx 0.01$ and the blue curve has $T/\mu \approx 0.006$. Both are at the critical lattice spacing $k_L = 1.27$. We see that the sum rule is satisfied after integrating up to a few times the UV scale $\mu$.  \label{fig:ReG} \label{fig:fPlot}}
\end{center}
\end{figure}

Let us examine the spectral weight transfer more closely. The conductivity sum rule implies that the integral of the real part of the conductivity over the positive real axis should not change upon introducing our ionic lattice. Therefore, at each order in the $\lambda$ expansion we can expect the correction to the conductivity to integrate to zero. This integral includes the contribution from the Drude peak delta function. The correction to the delta function can be read off from the zero frequency limit of the real part of the current current correlator. We can write
\be
\mathrm{Re}\,\sigma^{(2)} = -\pi \, \mathrm{Re}\,G^{(2)}_{J^yJ^y}(0)\,\delta\left(\omega\right) + \frac{\mathrm{Im}\,G^{(2)}_{J^yJ^y}\left(\omega\right)}{\omega} \,.
\ee
The left hand plot in figure \ref{fig:ReG} shows the real part of the current correlator at zero frequency as a function of temperature for two values of $\nu$. From this plot we see that there is delta function with negative weight in the correction to the conductivity. This in turn shows that there is spectral weight transfer from the Drude peak to the broad tail. The spectral weight transfer to the broad tail increases with $T$ in one case and decreases with $T$ in another. The right hand plot in figure \ref{fig:ReG} shows the integrated spectral weight
\begin{equation}\label{eq:f_definition}
f\left(\omega\right)=-\frac{\pi}{2} \mathrm{Re}\,G^{(2)}(0)+\int_{0^{+}}^{\omega}d\omega^{\prime}\,\frac{\mathrm{Im}\,G^{(2)}\left(\omega^{\prime}\right)}{\omega^{\prime}} \,,
\end{equation}
plotted at the critical momentum for the two different temperatures considered. The function asymptotes to zero, confirming the sum rule and providing a nontrivial check of our numerics. From the integrated spectral weight we see that the weight lost from the Drude peak already balances out the broad tail contribution, while the negative region we noted before in figure \ref{fig:OptCond} at higher frequencies is then balanced out by higher energy processes, and so seems not to be closely related to the critical power law tail.
Spectral weight redistribution from Drude peak to broad tail is a key experimental feature in unconventional metals, e.g. \cite{hussey, basov,boris, orenstein}.

Spectral weight redistribution has recently been discussed in the context of fermionic quantum BKT transitions \cite{pwp, pwp2}. There it has been suggested that the discrete scale invariance associated with the complex scaling dimensions in the unstable phase may be an underlying principle of the spectral weight redistribution occurring at Mott transitions.

In the remainder of this section we shall make some comments about the case in which there is an actual instability over some range of momenta.
A first question to address is whether spontaneous homogeneous currents are generated via coupling to the lattice. According to the analysis around the Reissner-Nordstr\"om black hole of \cite{Donos:2011bh}, the extremal near horizon version of which we repeated in \S \ref{sec:irspectrum}, we will have unstable modes at finite momenta. A perturbative analysis in $\lambda$ reveals that these modes will be unstable  around the backgrounds we are considering here as well. It was shown in \cite{Donos:2011bh} that a particular mode with wavenumber $k_{c}$ will first become unstable as we lower the temperature. Due to the presence of a lattice, we will now have the mixing of all modes with wavenumber $k_{c}+n\,k_{L}$ where $k_{L}$ is the lattice wavenumber and $n$ is integer. One of the fields that will participate in the unstable perturbation will be the gauge field. The question now is what would happen in the case where $k_{c}=m\,k_{L}$ for $m$ a positive integer. This is the critical case we are most interested in. Is a spontaneous homogeneous current generated? More generally, this question can be asked whenever $m \, k_L$ lies within the range of unstable momenta. We considered this case and explicitly constructed numerically the unstable mode at next to leading order in $\lambda$. Doing so revealed that the expectation value for the homogeneous current is zero to within our numerical accuracy. It seems that this result should be provable from general principles.

We have also studied the behavior of the conductivity close to the temperature where a striped phase transition occurs, assuming that our lattice is precisely such that $k_{c}=m\,k_{L}$ so that the unstable mode is directly communicated to the current. We find that the d.c.$\,$conductivity diverges as $\mathrm{Re}\left(\sigma^{(2)}\right)\propto \left(T-T_{c}^{\left(0\right)}\right)^{-2}$ where $T_{c}^{\left(0\right)}$ is specified by the existence of a static mode around the $\lambda=0$ background. The presence of a lattice will shift the critical temperature: $T_{c}=T_{c}^{\left(0\right)}+\lambda^{2}\,T_{c}^{\left(2\right)}+\cdots$ \cite{Flauger:2010tv}. The full correlator should diverge at $T_{c}$.

\section{Discussion}

In this discussion section we will reiterate the main points of our paper and simultaneously contextualize our results in terms of previous holographic and other theoretical work, highlighting potential experimental features.

\subsection{Broader perspective on linear resistivity, BKT and other models}
\label{sec:broader}

In \S \ref{sec:bkt} we described quantum BKT transitions from the standpoint of the near horizon geometry of an extremal black brane. In particular, the scaling dimension of the operator $\ocal$ took the form $\Delta = (2+z)/(2z) + \nu$, with $\nu$ the square root of a quantity that becomes negative beyond the critical point. This type of formula is typical in holographic frameworks. The fact that $2 \Delta = 1+2/z$ when $\nu=0$ underlies the scaling of the spectral weight at the critical point in (\ref{eq:critical}): $\lim_{\w \to 0} \text{Im} \, G_{\ocal \ocal}^R(\w)/\w \sim 1/T$. In the large $N$ limit, in which the holographic description is classical, this implies that the `double trace' operator $\ocal \ocal$ is marginal.

A more general field theoretical description of quantum BKT transitions \cite{Kaplan:2009kr} is as a merger and subsequent annihilation of two fixed points of the RG flow. At the transition point, where the two fixed points merge, one can expect in general that there should be a marginal operator (that does not exponentiate -- this statement at the critical point is in addition to the marginal operator that is presumably necessary to dial the dimensions of operators). It seems likely therefore that $\ocal \ocal$ will be marginal in general. However, away from the large $N$ limit, or some other kind of generalized mean field description, the dimensions of operators do not add and therefore we cannot draw any conclusions about the dimension of $\ocal$. It follows that $\ocal$ will not in general have the dimension $\Delta = (2+z)/(2z)$ at the critical point, ultimately not leading to a linear resistivity. A BKT transition in itself therefore does not appear to be sufficient. On the other hand, if the operator that condenses does admit a mean field description, then our mechanism for a linear resistivity will automatically go through. Theories with holographic duals are one set of concrete examples where this is the case, even when the formula for the operator dimension is more complicated than (\ref{eq:delta}). Other examples were given in \cite{Kaplan:2009kr}.
Note that ``mean field'' here does not imply that the transition is of mean field nature -- it is not -- rather that there exists a description in which the order parameter may be treated as a classical variable.

Beyond BKT transitions, previously studied models have used a conceptually similar approach of coupling the electric current to degrees of freedom with a $1/T$ spectral weight. For instance in \cite{kiv2,kiv3} such degrees of freedom resided on a density of impurities tuned to a specific value of the impurity coupling for which the model becomes solvable. The ``Gaussianization'' of the model at their critical coupling is reminiscent of our observation above that a mean field description of the unstable operator at the BKT transition is necessary to obtain the required temperature dependence of the spectral weight. The model also shares the feature of our approach that the Drude contribution to the conductivity is hoped to be subdominant relative to the quantum critical contribution.

A mean field limit was also tied to the emergence of a linear in temperature resistivity in Kondo lattice models, e.g. \cite{sy, georges, Sachdev:2010um, Sachdev:2010uj}. These models are not dissimilar to those mentioned in the previous paragraph. A $1/T$ spectral density of the localized spin excitations
is fed into the  self energy of the conduction fermions, leading to marginal Fermi liquid phenomenology \cite{Varma:1989zz}. Holographic realizations of similar models are mentioned in \S \ref{sec:holo} below, see \cite{Sachdev:2010uj} for an extended comparison.

In short, the general strategy is to find a well-motivated $1/T$ spectral density and then transfer this temperature dependence to the conductivity.
The holographic BKT transitions we have discussed provide a robust way to this, using irrelevant operators rather than explicit fermions to transfer the spectral weight, with the interesting behavior occurring at the boundary between a critical phase and an ordered phase.

\subsection{Experimental signatures of the mechanism}

Incorporating the effects of a Drude peak will presumably be necessary to connect with key features of the phase diagrams of unconventional materials, such as the crossover to Fermi liquid behavior away from the critical point. However, it seems possible to distill at least three generic ingredients of the scenario we have outlined that may manifest themselves experimentally in one way or another.

Firstly one could look for evidence of BKT scaling in quantities such as the critical temperature (\ref{eq:tc}) or the expectation value of the condensed operator. In general, on the side of the critical point in the phase diagram where an operator condenses, one expects quantities to be controlled by a dynamically generated scale with an exponential dependence on the distance to the critical point \cite{Kaplan:2009kr}.

A second generic feature is likely to be an asymmetry in the optical and d.c.$\,$conductivities, as well as other quantities that couple to the unstable mode, between the two sides of the quantum critical point. On the side where an operator condenses, at sufficiently low temperatures and frequencies, one expects only logarithmic corrections to the $1/T$ and $1/\w$ scalings of the critical point itself. See appendix \ref{eq:matching}. The other, disordered, side of the critical point should be characterized by varying power law dependences along the lines of equations (\ref{eq:fudged}) and (\ref{eq:sw}) above.

Thirdly, the nature of the transition, in which a scaling dimension becomes complex as two critical points annihilate \cite{Kaplan:2009kr}, requires the existence of a critical phase in the IR, of which the BKT quantum critical point is a boundary. This critical phase should manifest itself in scaling behaviors such as (\ref{eq:sw}) persisting down to arbitrarily low temperatures on the disordered side of the critical point. This scaling behavior may or may not require disentangling from other physics, such as superconductivity, present in the phase diagram.

\subsection{Context: linear resistivity in holography}
\label{sec:holo}

There exists a large body of interesting previous holographic work attempting to understand the origin of a linear in temperature d.c.$\,$resistivity. These works can broadly be classified into two classes, depending on the fate of the translation-invariance delta function in the conductivity. When the conductivity is due to probe branes \cite{Karch:2007pd, Hartnoll:2009ns} or probe fermions \cite{Faulkner:2010zz, Kachru:2010dk}, the degrees of freedom of interest are taken to be a parametrically small part of a bigger system. In this case the charge carriers can dump their momentum elsewhere and it does not return on the `experimental' timescale. A second option, that avoids diluting the carriers, is to explicitly break translation invariance via parametrically heavy degrees of freedom such as impurities \cite{Hartnoll:2008hs} or a lattice \cite{Hartnoll:2012rj, Horowitz:2012ky}.

For both classes of models, the d.c.$\,$ conductivity typically depends on parameters of the model. The dynamical critical exponent $z$ in \cite{Hartnoll:2009ns} and the dimension of IR operators in \cite{Faulkner:2010zz, Kachru:2010dk,Hartnoll:2008hs, Hartnoll:2012rj}. These models do not provide a compelling motivation, at least without additional input, for selecting the exponent that results in a linear resistivity. An interesting example of how additional input can help is \cite{Jensen:2011su}, where the dimension of the impurity operator of interest is protected by supersymmetry. It is furthermore unclear if the probe limit is a reasonable approximation for the real world systems of interest.

Our model has not addressed the fate of the delta function, but rather assumed that the corresponding Drude peak is swamped by a quantum critical contribution to the d.c. conductivity. We have suggested that this may be a reasonable starting point to discuss some bad metallic phases of matter. In contrast to the other models we find a linear in temperature resistivity occurring robustly and precisely where we would like it to: at a quantum critical point mediating the onset of an instability. The result is universal in the sense that it does not depend on the specific microscopic realization of the BKT transition. It does require, however, the existence of an irrelevant operator that can communicate the instability to the current correlators. We achieved such a coupling in \S \ref{sec:lattice} by having a vectorial instability at a finite wavenumber that was was communicated to the current via an irrelevant lattice.

\subsection{Open questions and future directions}

The essential feature of the holographic dynamics we have discussed is the BKT quantum phase transition.
The realization of such a BKT quantum phase transition in e.g. some Hubbard-like lattice model as a function of doping would of course be very exciting in bridging the gap between extremal horizons and the chemistry of unconventional materials.

Superconducting phases have been absent from our discussion thus far, and yet are a central feature of the most interesting bad metals. It seems reasonable to hope for a unified theory of bad metals and unconventional superconductivity. We can imagine two ways in which our mechanism could interplay with a superconducting instability. The first, a variant on a fairly conventional viewpoint, is simply that the superconducting instability could be triggered by coupling to the operator that develops the $1/T$ spectral weight at the critical point. A second possibility is that the critical phase itself, out of which the BKT transition emerges, is tightly entwined with
the existence of a superconducting instability. It is common in holographic frameworks, see \cite{Hartnoll:2012pp} for a recent discussion, for the IR of the superconducting phase to exhibit an emergent scaling symmetry. This could itself be the critical phase in which the BKT transition occurs. Such a picture would be, loosely speaking, in line with claims that the linear in T conductivity is associated to the existence of superconducting phases \cite{taillefer}. These critical issues deserve further thought.

Clearly a pressing issue is to establish (experimentally?) whether the picture we have outlined in figure \ref{fig:optical}, namely that in bad metals the Drude peak is swamped by an extended `critical' contribution, is reasonable in some circumstances. An alternative possibility is that the extended tail in fact falls rapidly to zero at $\w \approx 0$, so that for the strict d.c.$\,$conductivity the Drude contribution always dominates. This would incorporate naturally the fact that the d.c.$\,$conductivity is insensitive to the melting of an explicit Drude peak in the data.
Against this scenario we can note that, at least in the holographic models we have studied, the quantum critical contribution to the conductivity does extend all the way down to $\w=0$. An alternative possibility would be to build a model in which momentum relaxational process were built into the critical sector that the undergoes a BKT transition. In such a model we might hope to have our cake and eat it, simultaneously having a linear resistivity and incorporating the Drude peak physics.

Drude physics could explicitly be incorporated into our model by making the lattice potential (\ref{eq:AV}) periodic in both directions. One could imagine solving such a model numerically along the lines of \cite{Horowitz:2012ky}. Because the lattice is irrelevant in the IR, the momentum relaxation rate is very slow and hence one's first thought is that there will always necessarily be a well-defined Drude peak \cite{Hartnoll:2012rj}. However, sufficiently close to the BKT critical point, there is a further long timescale in the problem, the lifetime of the mode that is going unstable, and this may be able to transfer spectral weight out of the Drude peak and into the critical tail.
We made some preliminary remarks about this process in \S \ref{sec:numerics}. A more comprehensive holographic study of this point is desirable and has the potential to shed light on the nature of bad metals.

The particular model of a BKT transition that we constructed in \S \ref{sec:lattice} relied on local ($z=\infty$) criticality to obtain critical physics at nonzero wavenumber that could then be coupled by a lattice onto the homogeneous electrical conductivity. It would be very interesting to see if this requirement can be relaxed by combining for instance BKT transitions with Fermi surface physics, or by finding a way to communicate BKT transitions to the electrical current without going via finite wavevector modes.

The essential role of holography in our work has been to provide a setting where BKT transitions occur naturally and where the interplay between instability in a strongly correlated medium, lattices and conduction can be studied in a theoretically controlled way. Holography is less able to incorporate more conventional Fermi liquid physics in a useful way. Any strategy to describe the full phase diagram and the crossover the Fermi liquid behavior will presumably have to involve a mix of distilled insights from holography and conventional Fermi liquid computations.

Finally, the focus throughout this paper on the linear in temperature d.c.$\,$resistivity, and possible application to bad metals,
should not obscure the more general points we have made. Two further take home messages are firstly that strong optical and d.c.$\,$conductivities
can be achieved by coupling critical modes on the verge of an instability to the electrical current via irrelevant operators. Secondly, that (semi) local quantum criticality enables the effects of instabilities at nonzero wavevector to be efficiently coupled to e.g. the conductivity via lattice scattering.

\section*{Acknowledgements}

We have benefitted from discussions with Jerome Gauntlett, Chris Herzog, Steve Kivelson, John McGreevy, Mike Norman, Philip Phillips, Subir Sachdev, Dam Son and Jan Zaanen on topics related to those in this paper. We would also like to acknowledge the organizers of the Chicheley Hall meeting on ``Gravity, black holes and condensed matter'' for providing the stimulating setting where this collaboration was initiated. S.A.H. is partially funded by a Sloan research fellowship and by a DOE Early Career Award. A.D. is supported by an EPSRC fellowship.

\appendix

\section{Matching for low frequency and temperature correlators}
\label{eq:matching}

Consider a spacetime with an IR (finite temperature) scaling geometry and that is deformed by irrelevant operators to flow to an $AdS_4$ UV fixed point. Perturb the spacetime by a large number $N$ of fields $\Phi^I$. These fields are taken such that they are decoupled in the IR scaling geometry but generically all couple in the full spacetime. That is, they are coupled by the irrelevant operators participating in the RG flow described by the full background.

Solve the equations in the IR geometry, impose infalling boundary conditions, and then expand the solution near the boundary of the IR region (call it $\r \to 0$). The fact that the fields are decoupled in the IR implies that
\be\label{eq:IR}
\Phi^I(\rho) \sim c^I \Big( \rho^{1+2/z-\Delta_I} + \rho^{\Delta_I} \, {\mathcal G}^R_I(\w,T) \Big) \,.
\ee
Here $ {\mathcal G}^R_I(\w,T)$ is the IR Green's function, the $c^I$ are constants, and $\Delta_I$ is the IR dimension of the operator. This is equation (\ref{eq:GIR}) in the main text.

Away from the IR region, to leading order at low temperatures and frequencies we may set $\w = T = 0$. We assume that the fields satisfy second order equations in the radial direction. These may be, for instance, gauge invariant combinations of the bulk fields. Therefore there are $2N$ independent solutions. From their behavior near the UV boundary these break into $N$ non-normalizable and $N$ normalizable solutions. We wish to consider the response to a single source. Therefore we consider solutions of the form
\be
\Phi^I = A \Phi^I_\text{n.n.} + B \Phi^I_\text{n.} + \sum_{m = 1}^{N-1} a_m \Phi_m^I \,.
\ee
Here $\Phi^I_\text{n.n.}$ is the particular non-normalizable solution we wish to source, $\Phi^I_\text{n.}$ is the conjugate normalizable solution and the $\Phi_m^I$ are the remaining normalizable solutions. $A,B,a_m$ are numerical coefficients. The UV Green's function we wish to compute is
\be
G^R \propto \frac{B}{A} \,. 
\ee

The UV Green's function is now determined by a matching procedure given by a generalization of e.g. \cite{Faulkner:2011tm}. This is possible at low frequencies and temperatures. Expand the `far' solutions of the previous paragraph in the regime where they overlap with the near horizon solutions. One obtains
\bea
\Phi^I_\text{n.n.} & \sim & \a^I_- \, \rho^{1+2/z-\Delta_I} + \a^I_+ \, \rho^{\Delta_I} \,, \\
\Phi^I_\text{n} & \sim & \b^I_- \, \rho^{1+2/z-\Delta_I} + \b^I_+ \, \rho^{\Delta_I} \,, \\
\Phi^I_m & \sim & \g^I_{m-} \, \rho^{1+2/z-\Delta_I} + \g^I_{m+} \, \rho^{\Delta_I} \,. \label{eq:coefs}
\eea
Here the $\a_\pm^I, \b_\pm^I,\g^I_{m\pm}$ are numerical coefficients that can be determined in principle by solving the differential equations in the far region. They do not have any temperature or frequency dependence. Matching with the IR data (\ref{eq:IR}) now implies
\bea
c^I = A \, \a_+^I + B \, \b_+^I + \sum_m a_m \g^I_{m+} \,, \\
c^I {\mathcal G}^R_I(\w,T)= A\, \a_-^I + B \, \b_-^I + \sum_m a_m \g^I_{m-} \,.
\eea
There is no sum over $I$ in this equation. Taking an appropriate difference between these two equations gives
\be
A \, v^I + B \, w^I + \sum_m a_m z^{mI} = 0 \,,
\ee
where
\be
v^I = \a_+^I {\mathcal G}^R_I(\w,T) - \a_-^I \,, \quad w^I = \b_+^I {\mathcal G}^R_I(\w,T) - \b_-^I  \,, \quad z^{mI} = \g^I_{m+} {\mathcal G}^R_I(\w,T) - \g^I_{m-} \,.
\ee

The matrix $z^{mI}$ is an $(N-1)\times N$ matrix. It necessarily annihilates a specific vector $p^I$. The Green's function we are looking for can therefore be written
\be\label{eq:ba}
G^R \propto \frac{B}{A} \propto - \frac{v \cdot p}{w \cdot p} \,.
\ee
In practice, we can find $p$ by fixing one of its components and then determining the remainder by inverting a square submatrix of $z$. We see that the Green's function is a ratio of polynomials in ${\mathcal G}^R_I(\w,T)$.

We now wish to take the imaginary part of the Green's function. For real $\Delta_I$, the coefficients $\a_\pm^I, \b_\pm^I,\g^I_{m\pm}$ are real, whereas for complex $\Delta_I$, that is for $\nu_I$ pure imaginary, then the two coefficients appearing in each line of (\ref{eq:coefs}) are complex conjugates of each other. The case of real $\nu_I$ is simplest. In this case ${\mathcal G}^R_I \sim \text{max}(\w,T)^{2 \nu_I}$, which is small for temperatures and frequencies that are small compared to the UV scale. At a sufficiently generic point, in particular, away from instabilities that are not localized in the IR geometry, none of the $\a_\pm^I, \b_\pm^I,\g^I_{m\pm}$ vanish. Then one can simply expand the ratio of polynomials (\ref{eq:ba}) in the IR Green's functions. The result to leading nontrivial orders will have the form
\be
G^R(\w,T) = d^0 + \sum d^I \, {\mathcal G}^R_I(\w,T)  \,.
\ee
Because all the constants involved in this expression are real, taking the imaginary part we obtain
\be
\text{Im} \, G^R(\w,T)  = \sum d^I \, \text{Im} \, {\mathcal G}^R_I(\w,T) \,,
\ee
as quoted in the equation (\ref{eq:imgsum}) in the main text. In deriving this expression we could have allowed for the analytic $\w$ and $T$ dependence of the coefficients.

When some of the $\nu_I$ are imaginary, the Green's functions are not small and so we cannot expand the ratio of polynomials in (\ref{eq:ba}). In the d.c.$\,$limit $\w \ll T$ we can expand the IR Green's function in $\w$. Because, in the stable phase $T > T_c$, the full Green's function must vanish as $\w \to 0$, the result will take the form
\be
\text{Im} \, G^R(\w,T)  = \frac{\w}{T} F(\log T) \,,
\ee
for some in general complicated function $F$.

\end{document}